\documentclass[aps,pra,twocolumn,longbibliography]{revtex4-2}
\usepackage{amsmath,amsfonts,amsthm,bbm}
\usepackage[dvipsnames]{xcolor}
\usepackage{graphicx}
\usepackage{multirow,mhchem,chemformula}
\usepackage[colorlinks=true,breaklinks=true,linkcolor=red,citecolor=magenta,urlcolor=magenta]{hyperref}

\newcommand{\captiontitle}[1]{{\bf{#1.}}}
\newcommand{\sulfonium}{H$_3$S$^+$}
\newcommand{\error}[2]{#1$\pm$#2}
\newcommand{\phthrees}{Ph$_3$S$^+$}
\newcommand{\hartree}{$\mathrm{E_h}$ }
\newcommand{\device}[1]{$\mathsf{ibm\_#1}$}
\newcommand{\library}[1]{$\mathsf{#1}$}
\newcommand{\important}[1]{{\em{#1}}}
\newcommand{\cnot}{$\mathsf{cX}$ }
\newcommand{\Szero}{\mathrm{S}_0}
\newcommand{\Sone}{\mathrm{S}_1}
\newcommand{\Stwo}{\mathrm{S}_2}
\newcommand{\Tone}{\mathrm{T}_1}
\newcommand{\Ttwo}{\mathrm{T}_2}
\newcommand{\bgreek}[1]{{\boldsymbol{#1}}}
\newcommand{\vett}[1]{{\bf{#1}}}
\newcommand{\parm}{\bgreek{\theta}}
\newcommand{\angstrom}{\mathrm{\AA}}
\newcommand{\bs}{\vett{x}}
\newcommand{\crt}[1]{\hat{c}_{#1}^\dagger}
\newcommand{\dst}[1]{\hat{c}_{#1}^{\phantom{\dagger}}}
\newcommand{\oper}[1]{\hat{#1}}
\newcommand{\revision}[1]{{\color{black}#1}}

\begin{document}

\title{Quantum chemistry simulation of ground- and excited-state properties 
of the sulfonium cation on a superconducting quantum processor}

\author{Mario Motta}
\email{mario.motta@ibm.com}
\affiliation{IBM Quantum, IBM Research - Almaden, 650 Harry Road, San Jose, CA 95120, USA}
\author{Gavin O. Jones}
\email{gojones@us.ibm.com}
\affiliation{IBM Quantum, IBM Research - Almaden, 650 Harry Road, San Jose, CA 95120, USA}
\author{Julia E. Rice}
\affiliation{IBM Quantum, IBM Research - Almaden, 650 Harry Road, San Jose, CA 95120, USA}
\author{Tanvi P. Gujarati}
\affiliation{IBM Quantum, IBM Research - Almaden, 650 Harry Road, San Jose, CA 95120, USA}
\author{Rei Sakuma}
\affiliation{Materials Informatics Initiative, RD Technology \& Digital Transformation Center, JSR Corporation,\\
3-103-9, Tonomachi, Kawasaki-ku, Kawasaki, Kanagawa, 210-0821, Japan}
\author{Ieva Liepuoniute}
\affiliation{IBM Quantum, IBM Research - Almaden, 650 Harry Road, San Jose, CA 95120, USA}
\author{Jeannette M. Garcia}
 \email{jmgarcia@us.ibm.com}
\affiliation{IBM Quantum, IBM Research - Almaden, 650 Harry Road, San Jose, CA 95120, USA}
\author{Yu-ya Ohnishi}
 \email{yuuya\_oonishi@jsr.co.jp}
\affiliation{Materials Informatics Initiative, RD Technology \& Digital Transformation Center, JSR Corporation,\\
3-103-9, Tonomachi, Kawasaki-ku, Kawasaki, Kanagawa, 210-0821, Japan}

\begin{abstract}
The computational description of correlated electronic structure, 
and particularly of excited states of many-electron systems, 
is an anticipated application for quantum devices.
An important ramification is to determine the dominant 
molecular fragmentation pathways in photo-dissociation experiments 
of light-sensitive compounds, like sulfonium-based photo-acid 
generators used in photolithography.
Here we simulate the static and dynamical electronic 
structure of the \sulfonium molecule, taken as a minimal
model of a triply-bonded sulfur cation, on a 
superconducting quantum processor of the IBM Falcon architecture.
To this end, we generalize a qubit reduction technique termed
entanglement forging or EF [A. Eddins \textit{et al., Phys. Rev. X Quantum}, 2022, 
\textbf{3}, 010309], currently restricted to the evaluation of ground-state energies, 
to the treatment of molecular properties. While in a conventional quantum simulation 
a qubit represents a spin-orbital, within EF a qubit represents a spatial orbital,
reducing the number of required qubits by half. 
We combine the generalized EF with quantum subspace expansion [W. Colless et al, \textit{Phys. Rev. X}, 2018, \textbf{8}, 011021], a technique used to project the time-independent Schrodinger equation for ground- and excited-states in a subspace.
To enable experimental demonstration of this algorithmic workflow, we deploy 
a sequence of error-mitigation techniques.
We compute dipole structure factors and partial atomic charges along ground- and excited-state potential energy curves, revealing the occurrence of homo- and 
heterolytic fragmentation.
This study is an important step towards the computational description 
of photo-dissociation on near-term quantum devices, as it can be generalized
to other photodissociation processes and naturally extended in different ways to
achieve more realistic simulations.
\end{abstract}

\maketitle

\section{Introduction}

Solving the Schr\"{o}dinger equation for ground- and excited-states 
of many-electron quantum systems is one of the grand challenges of 
contemporary science \cite{helgaker2012recent,friesner2005ab}. 
In particular, the accurate computation of excited-state properties by 
numerical simulations stands to impact many problems in pure and 
applied quantum chemistry, exemplified by photochemical processes 
that result from absorption of photons
and promotion of electrons to excited states.

The semiconductor industry has employed these processes to use
photolithographic materials in solid-state chip fabrication \cite{gates2005new,gangnaik2017new}. Fundamentally, 
fabrication involves coating a silicon wafer with a 
thin film of a precisely engineered block co-polymer with different 
functional side chains on blocks that self-assemble into lamellae when processed. 
In this example, blocks have distinct reaction profiles, and some 
engineered systems include acid-sensitive side chains that, when 
reacted, change the block solubility coefficient. 
A photo-acid generator (PAG) can be embedded in the polymer film and 
photochemically reacts at specific wavelengths of light to release 
a free proton in the solid-state that can subsequently react with
acid-sensitive side chains \cite{nalamasu1990development,fallica2018photoacid,martin2018recent,sambath2020bodipy}. 
Patterns form with the use of a mask that blocks or exposes different 
parts of the film to ultraviolet (UV) light, modifying the solubility 
of the polymer such that it can be selectively washed away from the 
wafer in aqueous solvents. One such effective industrial PAG contains 
the triphenylsulfonium (\phthrees) cation \cite{ohmori1998ab,dektar1988triphenylsulfonium,dektar1990photochemistry,klikovits2017novel,jin2014pi,zhou2002efficient}.

The computational description of these photochemical processes poses 
a number of challenges, for example: characterizing the light-matter 
interaction to determine transitions to electronic (and, in general, vibronic) 
excited states; and, in photo-dissociation reactions, 
assessing the electronic structure of excited states to determine the 
nature of the dissociation path (e.g. homolytic \cite{knapczyk1969reactions,crivello1979dye,pappas1984photoinitiation} 
or heterolytic \cite{davidson1982some}).
In addition, qualitatively correct and quantitatively accurate 
calculations require incorporating solvation and thermal effects, and 
reliably assessing the electronic structure of the studied species by
accounting for static and dynamical correlations in realistic basis sets.
Capturing such effects accurately
is essential for understanding molecular properties, for predictive computations, 
and ultimately for introducing new PAGs, because the current semiconductor 
process requires molecular-size order controlling.

Over the last decades, research in computational many-electron quantum 
mechanics has generated algorithms for conventional classical computers
that yield approximate, though often very accurate, estimates of ground- 
and excited-state molecular properties at polynomial cost  \cite{LeBlanc_PRX_2015,Zheng_Science_2017,Motta_PRX_2017,williams2020direct}. 

Digital quantum computers are an alternative and complementary framework 
to simulate many-body quantum systems
\cite{georgescu2014quantum,cao2019quantum,bauer2020quantum,motta2021emerging}. 
Assuming high-quality qubits in a sufficiently large number,
they allow simulation of the time-dependent Schr\"{o}dinger equation at 
polynomial cost introducing controllable approximations only 
\cite{lloyd1996universal,martyn2021grand}, 
thus being capable of accessing a vast class of excited-state properties.
Recent advances in hardware manufacturing has produced quantum computers 
that can carry out computations on a limited scale. 
Despite the rapid development of quantum hardware, modern quantum 
computation platforms are immature. As a consequence, simulations of excited
states on near-term devices are typically restricted to heuristic quantum
subspace algorithms \cite{mcclean2017subspace,takeshita2020subspace,cohn2021filterdiagonalization,yoshioka2021virtualsubspace,epperly2021subspacediagonalization,baek2022say,colless2018computation}, that yield approximations to excited-state 
wavefunctions and properties within the budget of these devices by projecting
the Schr\"{o}dinger equation onto a suitably constructed subspace.
It is therefore a real possibility, and of central importance at this time, 
to assess the potential usefulness of near-term quantum devices on problems 
of conceptual and practical interest, e.g. the computation of molecular 
excited states.

Here, we report the development of a heuristic methodology 
that leverages structured entanglement in many-electron wavefunctions to 
calculate ground- and excited-state molecular properties, and its experimental demonstration
on a superconducting quantum processor.
More specifically, we generalize a qubit reduction technique called entanglement
forging (EF) \cite{eddins2021doubling}, initially proposed for variational simulations of ground-state energies, to the computation of generic many-body observables. 
While in a conventional quantum simulation 
a qubit represents a spin-orbital, within EF a qubit represents a spatial orbital,
reducing the number of required qubits by half. 

To improve the accuracy of this technique, and to approximate excited-state energies and
properties, we combine EF with quantum subspace expansion (QSE), an example of a heuristic quantum subspace algorithm \cite{mcclean2017subspace,colless2018computation,smart2021quantum}
which, in its simplest form, projects the Schr\"{o}dinger equation onto a subspace spanned by single and double excitations on top of a reference wavefunction.
The proposed methodology extends the applicability of EF, 
allowing the computation of a significant set of observables, 
and that of QSE, 
facilitating its demonstration on contemporary quantum hardware 
due to the qubit reduction operated by EF.

We apply the proposed technique, in combination with multiple error mitigation methods, 
to investigate the gas-phase photo-dissociation of \sulfonium, taken as the simplest molecular model for \phthrees.
Common to both compounds is the presence of a triply-bonded sulfur cation. The most accurate description of the computational model for \phthrees requires the inclusion of the $\pi$-conjugated phenyl groups, which determines the energy and character of its excited states, but this feature corresponds to a higher computational cost which makes \sulfonium a more suitable target for simulations on contemporary quantum hardware.

We assess the interaction between \sulfonium and UV light within the electric 
dipole approximation in response theory, and characterize dissociation paths as
homo- or heterolytic by computing partial atomic charges and other properties 
of the excited-states. 
Our study contains approximations and limitations, that we endeavor to document. Notwithstanding these limitations, it illustrates that near-term quantum hardware can be effectively used to explore ground- and excited-states of molecules by means of active-space calculations. While active spaces treated in our study are still small, the underlying methodology naturally extends to larger active spaces. Furthermore, our algorithm is amenable to multiple algorithmic improvements and extensions
(for example to treat dynamical correlation, solvation effects, and larger chemical systems), that draw a path towards larger and more realistic simulations.

\section{Methods}
\label{sec:methods}

Several authors have shown that the absorption cross-section of electromagnetic
radiation by a molecular system can very generally be represented as a Fourier integral \cite{lax1952franck,heller1978photofragmentation,kulander1978time,johnson1989recurrences}.

Let $\oper{H}$ be the  unperturbed time-independent molecular Hamiltonian, with eigenstates
$\oper{H} | \Phi_A \rangle = \varepsilon_A | \Phi_A \rangle$. If the system, initially
at equilibrium at zero temperature, interacts weakly with an external electric field of frequency $\omega$, transitions from the ground state  into other quantum states $| \Phi_A \rangle$ occur if the frequency of the radiation is close to $\Delta \varepsilon_{A0} = \varepsilon_A - \varepsilon_0$. Assuming a field with wavelength much larger than 
molecular dimensions, the perturbation can be written as $\oper{V}(t) = - \hat{{\bgreek{\mu}}} \cdot \hat{{\bgreek{E}}}(t)$, where $\hat{{\bgreek{\mu}}}$ is the dipole moment operator. 
According to time-dependent quantum-mechanical perturbation theory, to first order in the perturbation, the rate of transition from the ground state to 
any excited state is given by Fermi's golden rule \cite{gordon1965molecular,boulet1982short,clerk2010introduction,vitale2015anharmonic,nascimento2016linear,goings2018real,li2020real} and is proportional to the
dipole dynamical structure factor (DSF),
\begin{equation}
\label{eq:structure_factor}
S(\omega) 
= \int \frac{dt}{2\pi} e^{i\omega t} 
\langle \Phi_0 | \hat{{\bgreek{\mu}}}(t) \cdot \hat{{\bgreek{\mu}}} | \Phi_0 \rangle 
=
\sum_A \mu_{A0} \, \delta(\omega - \Delta \varepsilon_{A0})
\;,
\end{equation}
where $\mu_{A0} = | \langle \Phi_A | \hat{{\bgreek{\mu}}} | \Phi_0 \rangle |^2$ is the transition dipole between Hamiltonian eigenstates $\Phi_0$ and $\Phi_A$. Combined with the excitation energy $\Delta \varepsilon_{A0}$, it permits evaluation of the oscillator strength 
$\mu_{A0} \Delta \varepsilon_{A0}$, which in turn specifies the 
absorption cross section of electromagnetic radiation \cite{vitale2015anharmonic,nascimento2016linear,goings2018real,li2020real}.
It should be noted that the scalar character of the DSF is due to the assumption of an isotropic system, for which any response is independent 
from the polarization vector of the incident radiation. 
Furthermore, while an appropriate description of photo-dissociation requires
a joint quantum mechanical treatment of electrons and nuclei \cite{gordon1965molecular,boulet1982short}, particularly near conical intersections, here, concerned with valence-electron UV/vis spectroscopy, we focus on vertical 
electronic transitions.

\begin{figure}[t!]
\includegraphics[width=0.85\columnwidth]{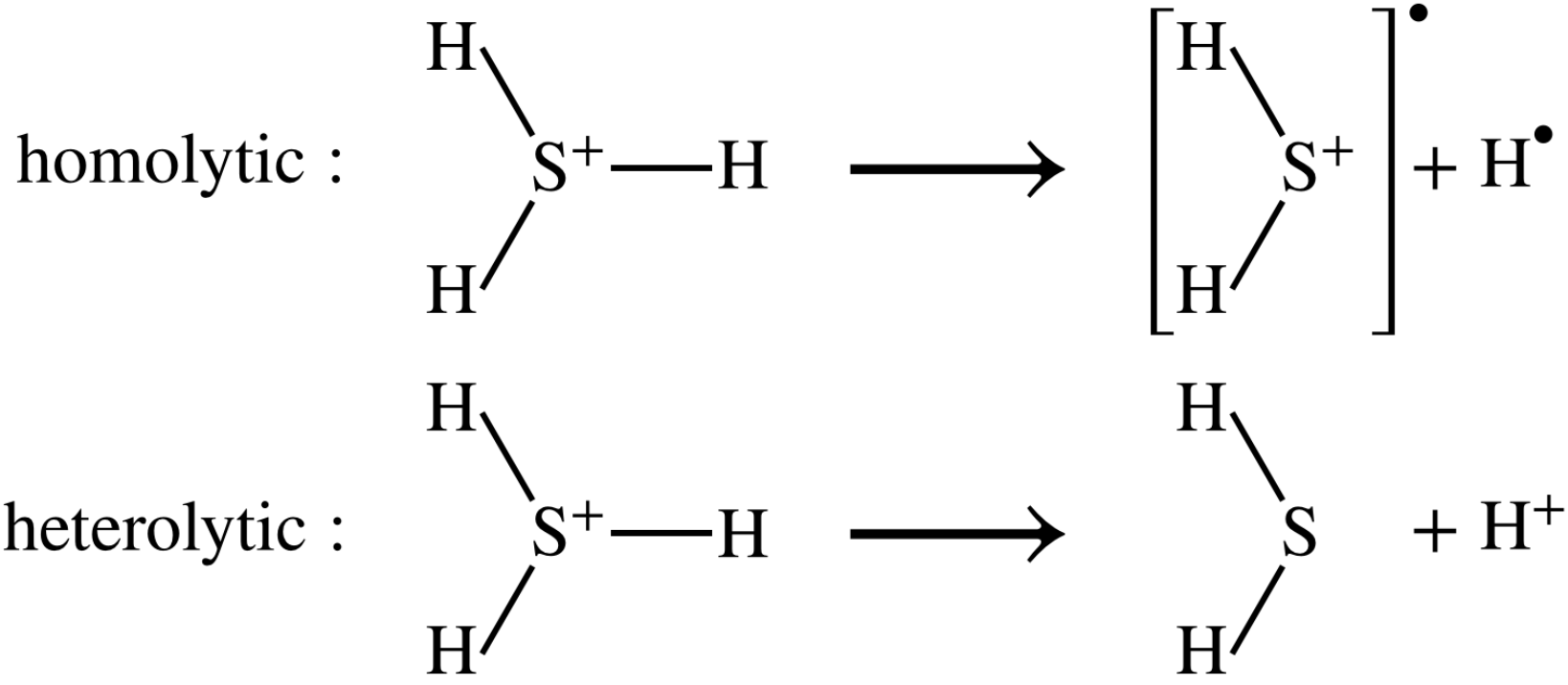}
\caption{\captiontitle{Molecular fragmentation paths} In the homolytic 
cleavage of a single SH bond of \sulfonium, the two electrons in the 
bond are divided equally between $\ce{H2S}$ and $\ce{H}$, leading to 
the formation of two radicals (top, marked by black circles). 
In the heterolytic cleavage, the two electrons are taken by one part 
of the bond, with formation of closed-shell products $\ce{H2S}$ and $\ce{H^+}$.
}
\label{figure:FIG1}
\end{figure}

DSFs of the form shown in Eq.~\eqref{eq:structure_factor} are natural
targets for quantum computers, where the time-dependent dipole correlation 
function $f(t) = \langle \Phi_0 | \hat{{\bgreek{\mu}}}(t) \cdot \hat{{\bgreek{\mu}}} | \Phi_0 \rangle$ can be computed by simulating Hamiltonian time evolution \cite{motta2021emerging}.
Note that $f(t)$ needs to be computed over a sufficiently long time interval to allow for accurate reconstruction of its Fourier \revision{transform}, time evolution 
needs to be controllably approximated e.g. with product formulas, and computation of correlation functions 
involves deep quantum circuits comprising the Hadamard 
test or mid-circuit measurements \cite{endo2020calculation}.
Approximate ground- and excited states can also be computed 
by means of quantum diagonalization algorithms.
These heuristic methods may require shallower quantum circuits, making them compatible with near-term quantum computers. Furthermore, they enable evaluation of Eq.~\eqref{eq:structure_factor} directly in the
frequency domain; \revision{for example, they allow to evaluate the right member of Eq.~\eqref{eq:structure_factor} by summing over the set of computed excited states without computing time-dependent correlation functions}.
Access to excited-state wavefunctions also allows computing densities
$\rho_A({\bf{x}}) =\langle \Phi_A | \sum_\sigma \crt{\sigma}({\bf{x}}) 
\dst{\sigma}({\bf{x}}) | \Phi_A \rangle$, where the operator 
$\crt{\sigma}({\bf{x}})/\dst{\sigma}({\bf{x}})$ creates or destroys an
electron with spin $\sigma$ at position ${\bf{x}}$.
From these, one can extract partial atomic charges, which in turn allow characterization of 
the dissociation of a single SH bond as a homolytic ($\ce{H3S+ \to H2S+ + H}$) 
or heterolytic ($\ce{H3S+ \to H2S + H+}$) process, as illustrated in Figure
\ref{figure:FIG1}.

\subsection{Algorithm.}

In this Section, we describe the algorithms used in the present work. 
Additional methodological details are provided in Appendix \ref{sec:app_a}-\ref{sec:app_c}.

\subsubsection{Classical preprocessing.} The starting point of this study is 
the definition of a Hamiltonian operator $\hat{H}$. 
To elucidate the electronic structure of \sulfonium along cleavage of a 
single SH bond, we performed a set of constrained geometry optimizations 
on a classical computer and, for each geometry, projected the electronic 
Hamiltonian onto an active space of 6 spatial electrons corresponding to
sulfur 3p and hydrogen 1s. 
Although the active-space approximation biases the electronic structure 
of \sulfonium, it offers the possibility to benchmark the performance of 
heuristic quantum algorithms and near-term quantum hardware, 
while establishing a foundation for scaling to more complex quantum 
simulations.
Additional details are provided in Appendix \ref{sec:app_a}.

\subsubsection{Ground-state calculations.} 
Having defined a Hamiltonian, we start the search for excited states with
a preliminary ground-state calculation. 
To this end, we resort to the EF technique \cite{bravyi2016trading,eddins2021doubling,huembeli2022entanglement},
based on the idea of partitioning a register of qubits in two halves, and 
representing the target wavefunction as 
$| \Psi_{\parm} \rangle = \sum_k \lambda_k \oper{U}(\parm) | \bs_k \rangle \otimes \oper{U}(\parm) | \bs_k \rangle$,
where $\oper{U}(\parm)$ is a parameterized unitary, $\lambda_k$ are a set 
of coefficients, and $| \bs_k \rangle$ are a set of computational basis states. 
Observables are written as linear combinations of tensor products 
$\oper{A} \otimes \oper{B}$, and their expectation values are expressed \revision{(see Appendix~\ref{sec:app_b} for a derivation)} as
\begin{equation}
\label{eq:ef_1}
\langle \Psi_{\parm} | \oper{A} \otimes \oper{B} | \Psi_{\parm} \rangle 
= 
\sum_{kl} \lambda_k \lambda_l \, A_{kl} B_{kl} 
\;,
\end{equation}
where $X_{kl} = \langle \bs_k | \oper{U}^\dagger (\parm) \oper{X} \oper{U}(\parm) | \bs_l \rangle$
is evaluated as
\begin{equation}
\label{eq:ef_2}
X_{kl} 
= \sum_{p=0}^3 \frac{(-i)^p}{4}
\langle \phi^p_{kl} | \oper{X} | \phi^p_{kl} \rangle
\,,\,
| \phi^p_{kl} \rangle 
= \frac{ |{\bf{x}}_k \rangle + i^p | {\bf{x}}_l \rangle }{ \sqrt{2} }
\,.
\end{equation}

\begin{figure*}[t!]
\includegraphics[width=\textwidth]{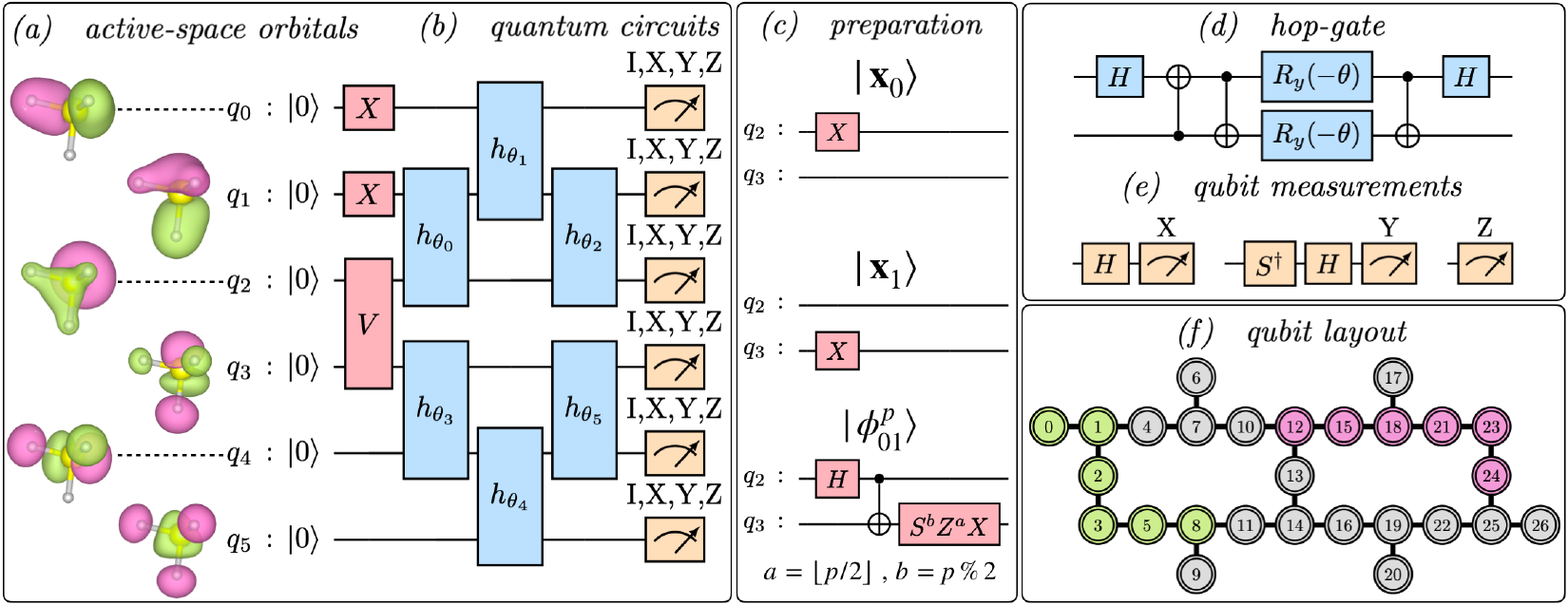}
\caption{ \captiontitle{Qubit layout and quantum circuits} 
$(a)$ active-space orbitals 
of \sulfonium at equilibrium geometry, obtained from a mean-field simulation of the 
Born-Oppenheimer Hamiltonian, and mapped onto qubits as illustrated. 
$(b)$ the circuit diagram depicts an entanglement forging Ansatz, comprising 
an initial state preparation (light red box) followed by 3 layers of parametrized 2-qubit 
hop-gates in a brickwall pattern (light blue box) and a final measurement (light orange 
meter symbols). $(c)$ the red box marked $V$ in panel $(b)$ completes the initialization 
of 6 qubits in a computational basis state ${\bf{x}}_k \in \{ |111000 \rangle, |110100 \rangle \}$ 
(top, middle circuits marked ${\bf{x}}_0$, ${\bf{x}}_1$ respectively) 
or in a superposition state 
$|\phi^p_{01} \rangle = \left( |{\bf{x}}_0 \rangle + i^p | {\bf{x}}_1 \rangle \right) / \sqrt{2}$ 
(bottom circuit marked $\phi^p_{01}$, where $a=\lfloor p/2 \rfloor = 0,0,1,1$ and $b= p\% 2 = 0,1,0,1$ for $p=0,1,2,3$ respectively).
$(d)$ compilation of a 2-qubit hop-gate into single-qubit and \cnot gates.
$(e)$ depiction of the final measurement operations.
$(f)$ depiction of two 6-qubit lines on sub-grids of the \device{kolkata} device. 
Each circuit involves up to 19 \cnot gates, 42 single-qubit gates, and 8 variational parameters,
and only requires gates between pairs of qubits adjacent in a linear topology.
}
\label{figure:FIG2}
\end{figure*}

Within EF, one prepares the states $| \phi^p_{kl} \rangle$ on a quantum processor, 
measures the matrix elements $X_{kl}$ for $\oper{X} = \oper{A},\oper{B}$ and the 
expectation values in Eq.~\eqref{eq:ef_1}.
In this formalism, a qubit represents a spatial orbital rather than a spin-orbital, and thus the number of qubits required for a simulation is reduced by half. 

The unitary $\oper{U}(\parm)$ and states $| \bs_k \rangle$ are Ans\"{a}tze, 
and parameters $\parm$ are optimized variationally along with coefficients $\lambda_k$. 
Here, we choose ${\bf{x}}_k \in \{ |111000 \rangle, |110100 \rangle \}$ to highlight 
entanglement across frontier active-space orbitals, and
$\oper{U}(\parm)$ as a product of 6 ``hop-gates" (i.e. number-conserving 
functionally complete 2-qubit gates). The circuits executed in this work 
are shown in Fig.~\ref{figure:FIG2}$a$-$e$, and additional details are in Appendix~\ref{sec:app_b}.

\subsubsection{Excited-state calculations.} 
To access excited states, we extend EF to encompass the framework of 
quantum diagonalization algorithms, exemplified by QSE \cite{mcclean2017subspace}.
Within QSE, a set of excitation operators $\{ \oper{E}_\mu \}_\mu$ are 
chosen, Hamiltonian and metric matrices are constructed as
$H_{\mu\nu} = \langle \Psi_{\parm} | \oper{E}_\mu^\dagger \oper{H} \oper{E}_\nu | \Psi_{\parm} \rangle$
and
$M_{\mu\nu} = \langle \Psi_{\parm} | \oper{E}_\mu^\dagger \oper{E}_\nu | \Psi_{\parm} \rangle$ 
respectively, and Hamiltonian eigenstates are approximated as 
$| \Phi_A \rangle = \sum_\nu c_{\nu A} \oper{E}_\nu | \Psi_{\parm} \rangle$, 
where the columns of $c$ are solutions of the eigenvalue equation 
$H c_A = M c_A \varepsilon_A$. 
Here, we employ as excitation operators single- and double-electronic excitations, 
as this choice is natural for electronic systems and compatible 
with the representation of QSE matrices by Eq.~\eqref{eq:ef_2}.

\subsubsection{Evaluation of observables.}
The workflow outlined here allows access to ground-, 
excited-state, and transition matrix elements of a vast class of operators.
Along with the Hamiltonian, we compute the electron number and total spin 
operators, respectively $\hat{N}$ and $\hat{S}^2$, two important constants 
of motion used to label excited states and classify transitions, and the 
density matrices (RDMs) $\rho^{AB}_{pr} = \langle \Phi_A | \sum_\sigma 
\crt{p \sigma} \dst{r \sigma} | \Phi_B \rangle$.
Transition RDMs $\rho^{A0}$ provide access to the dipole DSF
Eq.~\eqref{eq:structure_factor}, whereas ground- and excited-state RDMs 
(respectively $\rho^{00}$ and $\rho^{AA}$) provide access to partial atomic
charges.

\subsection{Hardware experiments}

Simulations were run on IBM's 27-qubit processor \device{kolkata} based 
on the Falcon architecture, as shown in Figure \ref{figure:FIG2}$f$.
Segments of best-performing qubits were selected monitoring average readout 
and \cnot errors, and IBM's \library{Qiskit} and \library{runtime} 
libraries were used to interface with quantum hardware \cite{aleksandrowicz2019qiskit}.
Along with hardware simulations, we performed noiseless and noisy 
simulations of quantum circuits using the \library{statevector} and 
\library{qasm} simulators of \library{Qiskit}, and exact diagonalization 
(full configuration interaction of FCI) calculations using 
\library{PySCF} \cite{sun2018pyscf,sun2020recent}.
To reduce decoherence effects and systematic errors occurring on quantum 
hardware, we resorted to a combination of error mitigation techniques, 
detailed below and further discussed in Appendix \ref{sec:app_c}.

\subsubsection{Readout error mitigation} \label{par:roem}
In general, measurement errors over $n$ qubits satisfy the relation $A {\bf{p}}_{ideal} = {\bf{p}}_{noisy}$ where ${\bf{p}}_{noisy}$ and
${\bf{p}}_{ideal}$ are vectors of probabilities (the former is returned by the noisy quantum system and the latter contains probabilities 
in the absence of measurement errors) and $A$ is a $2^n \times 2^n$ 
complete assignment matrix.
To mitigate readout errors in such scenario, one needs to execute $2^n$ circuits to measure ${\bf{p}}_{noisy}$ and compute $A$, and solve a
square system of $2^n$ linear equations to compute ${\bf{p}}_{ideal}$ \cite{maciejewski2020mitigation}.
However, it is often the case that errors on multiple qubits can be well approximated using at most $\mathcal{O}(n)$ calibration circuits \cite{nation2021scalable}. This result holds when: $A$ can be approximated
as a tensor product of $n$ matrices of shape $2 \times 2$ (tensored Ansatz); $A$ is diagonally-dominated, allowing efficient matrix-free solution of 
the linear system; ${\bf{p}}_{noisy}$ is well-approximated by a vector 
with sparsity at most $\mathcal{O}(n_s)$, where $n_s$ is the number of accumulated statistical samples (or shots).

\begin{figure*}[t!]
\centering
\includegraphics[width=0.9\textwidth]{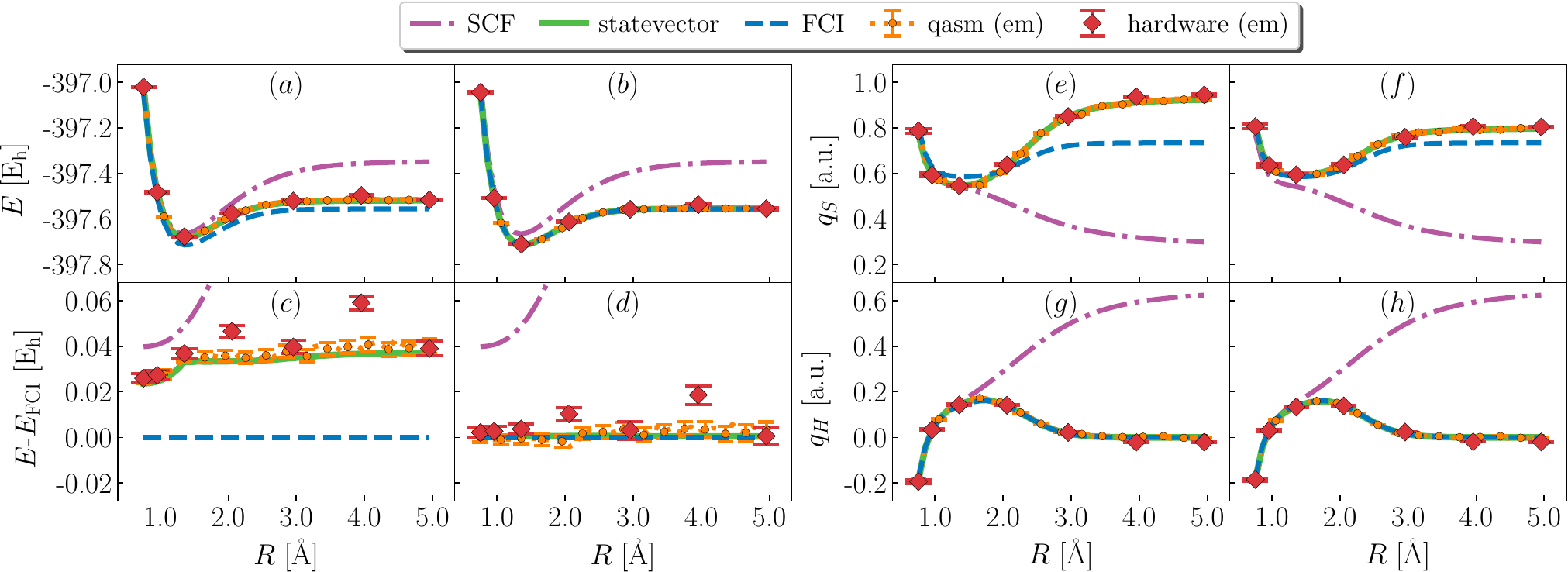}
\caption{ \captiontitle{Ground-state energies and partial atomic charges 
during bond cleavage} Left: computed total energy ($a,b$) and deviation between 
computed and FCI total energy ($c,d$) from EF ($a,c$) and QSE with single and 
double excitations on top of EF ($b,d$) using simulators (green lines, orange symbols 
for \library{statevector} and \library{qasm}) and quantum hardware 
(\device{kolkata}, red symbols). \important{em} is an abbreviations for error 
mitigation. Right: computed atomic charges on S ($e,f$) and the departing H 
($g,h$) as a function of bond-length from EF ($e,g$) and QSE ($f,h$).
Charges are computed with a Mulliken population analysis based on meta-Lowdin 
atomic orbitals. Partial atomic charges on the remaining H atoms are equal to 
each other,and to $(1-q_S-q_H)/2$.
}
\label{figure:FIG3}
\end{figure*}

\begin{figure*}[t!]
\centering
\includegraphics[width=0.55\textwidth]{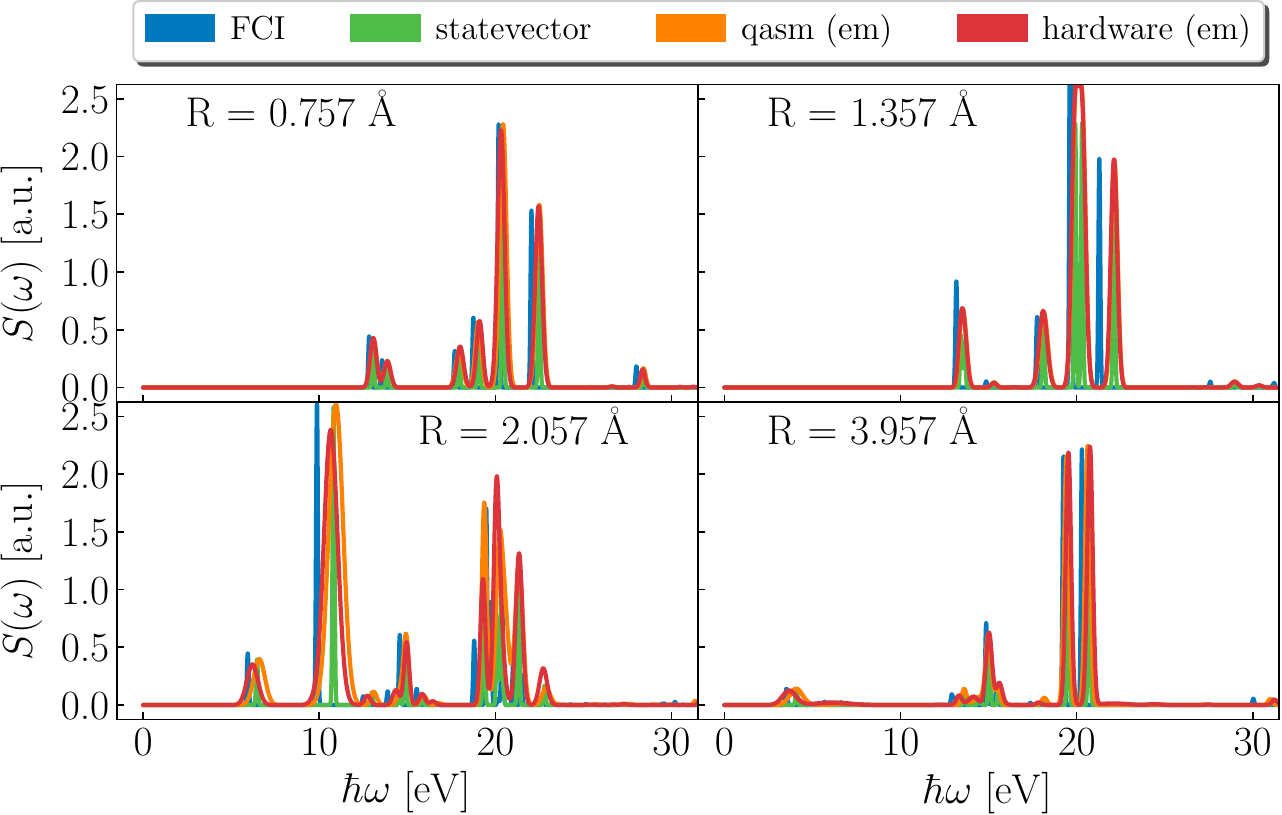}
\caption{\captiontitle{Dipole spectral functions}
Spectral function of the dipole operator from FCI (blue), \library{statevector} (green), 
\library{qasm} with noise model and error mitigation (orange) and hardware (\device{kolkata}, red)
at the representative bond-lengths $R=$ 0.757, 1.357, 2.057 and 3.957 $\angstrom$ (left to right). 
Spectral peaks are plotted with a broadening of 0.2 mHa (5 meV) for FCI and \library{statevector}, 
and a broadening reflecting the uncertainty on the excitation energy for \library{qasm} and \device{kolkata}. 
}
\label{figure:FIG4}
\end{figure*}

\subsubsection{Post-selection} Since the states $| \phi^p_{kl} \rangle$ have $3$ electrons, 
only outcomes of a computational basis measurement corresponding to binary strings with 
Hamming weight 3 are retained \cite{huggins2021efficient}.
Furthermore, $| \phi^p_{kl} \rangle$ is 
real-valued for $p=0,2$ and thus expectation 
values of purely-imaginary Pauli operators on 
such states vanish.

\subsubsection{Clifford-based gate error mitigation}
For particular parameter configurations, e.g. $\parm^* = \vett{0}$, 
the circuits in Fig.~\ref{figure:FIG2}$b$-$e$ are in the Clifford group.
The expectation value of a linear combination of Pauli operators over such 
a circuit can be computed exactly at polynomial cost on a classical computer 
\cite{gottesman1998heisenberg} and measured on a device,
offering a pool of data from which to learn the effect of decoherence
on measurement outcomes, and mitigate errors \cite{czarnik2021error}.
We use such data to perform an add-and-subtract correction, i.e. 
to compute 
\begin{equation}
X_{\mathrm{ideal}}(\parm) \simeq X_{\mathrm{hw}}(\parm) + X_{\mathrm{ideal}}(\parm^*) - X_{\mathrm{hw}}(\parm^*) \quad.
\end{equation}

\subsubsection{Purification} \label{par:pur} Any $n$-qubit density operator can be written as 
\begin{equation}
\oper{\rho} = 2^{-n} \Big( \oper{\mathbbm{1}} + \sum_i a_i \oper{\sigma}_i \Big) \quad,\quad \vec a \in \mathbbm{R}^{4^n-1} \quad,
\end{equation}
where the Bloch vector $\vec a$ is defined so that $\oper{\rho} = \oper{\rho}^\dagger$ and $\mbox{Tr}[\oper{\rho}]=1$, and must be compatible with the condition $\oper{\rho} \geq 0$.
In particular, since the purity $\mbox{Tr}[\oper{\rho}^2]$ of a density operator lies \cite{nielsen2002quantum} in the interval $[2^{-n},1]$, and $\mbox{Tr}[\oper{\rho}^2] = 2^{-n}(1+\| \vec a \|^2)$, then $\| \vec a \|^2$ must lie in the interval $[0,2^n-1]$.
Due to decoherence and artifacts of error mitigation, we may observe $\mbox{Tr}[\oper{\rho}^2] \notin [2^{-n},1]$ despite the target state being pure. 
When that happens, we scale the Bloch vector so that $\mbox{Tr}[\oper{\rho}^2] = 1$.
In the remainder of this work raw, readout-error mitigated (only technique $a$), and fully 
error-mitigated results (\revision{all of the four mitigation techniques in subsections 
\ref{par:roem} to \ref{par:pur}}) will be labeled \important{raw}, 
\important{roem}, and \important{em} respectively.

\begin{figure*}[t!]
\centering
\includegraphics[width=0.8\textwidth]{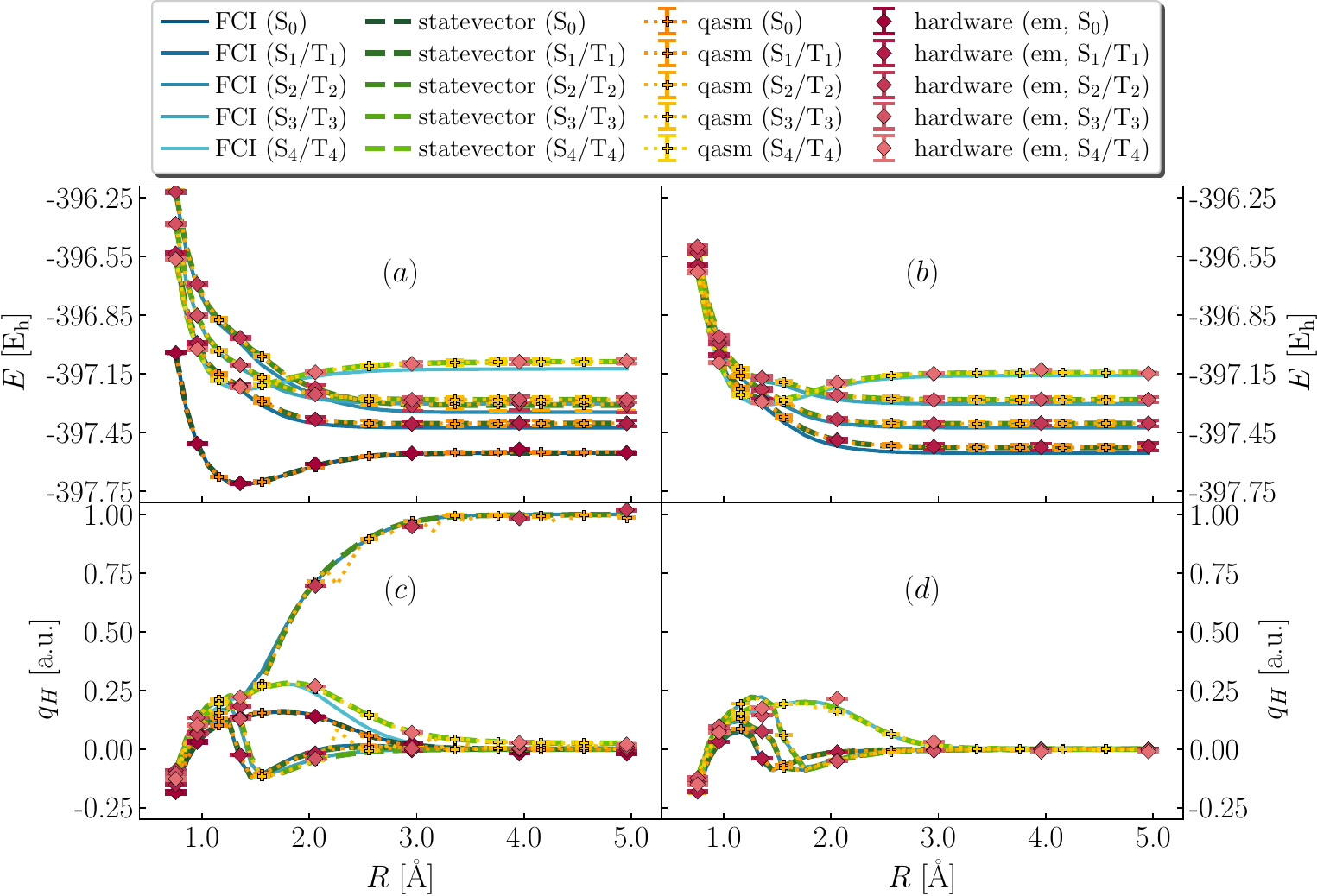}
\caption{ \captiontitle{Excited-state energies and partial atomic charges during bond cleavage}
\revision{
Panels $a$ and $b$: total energies of four low-lying excited singlet (left, panel $a$) and  triplet (right, panel $b$) states from FCI (blue lines), simulators (green lines, orange symbols for \library{statevector}, \library{qasm}) and quantum hardware (\device{kolkata}, red symbols).
Panels $c$ and $d$: computed partial atomic charges on the departing H as a function of SH bond-length for the singlet (left, panel $c$) and triplet (right, panel $d$) excited  states, from QSE with singles and doubles on top of the EF wavefunction, using FCI (blue lines), simulators (green lines, orange symbols for \library{statevector}, \library{qasm}), and quantum hardware (\device{kolkata}, red symbols). 
Dark and light colors indicate the lower- and higher-energy states in the large bond-length regime $R$ respectively.}
}
\label{figure:FIG5}
\end{figure*}

\section{Results}
\label{sec:results}

\subsection{Ground-state simulations}

Figure~\ref{figure:FIG3} shows simulation results for the ground state of \sulfonium. 
Variational EF simulations (panels $a,c$) are on average $\sim 40$ m\hartree above FCI,
with a non-parallelity error (defined as 
$\mathsf{npe} = \max_R | \Delta E(R) - \langle \Delta E(R) \rangle |$ 
with $\Delta E(R) = E(R) - E_{\mathrm{FCI}}(R)$) of 10.2, \error{11}{3}, \error{20}{3} 
m\hartree and a binding energy (defined as $E(R_{\mathrm{max}}) - \min_R E(R)$ of 163, 
\error{164}{4}, \error{162}{4} m\hartree for \library{statevector}, \library{qasm} and 
\device{kolkata} respectively.
It should be noted that binding energies are challenging to evaluate on quantum hardware,
due to a non-smooth potential energy curve for large $R$.
The combined use of EF and QSE (panels $b,d$) 
significantly improves the agreement between 
variational and FCI energies, and decreases
non-parallelity errors to 0.4, 
\error{2}{6}, \error{6}{13} m\hartree for \library{statevector}, \library{qasm} and 
\device{kolkata} respectively.
Binding energies decrease slightly, respectively 
to 159.0, \error{163}{4}, \error{156}{5}
 m\hartree.

Figure~\ref{figure:FIG3} also shows ground-state partial atomic charges. 
Hartree-Fock (SCF) incorrectly predicts (panels $g,h$) the charge $q_H$ on the departing 
hydrogen to remain finite as $R$ diverges, i.e. a heterolytic ground-state dissociation path.
All other methods predict $q_H$ to vanish as $R$ diverges, i.e. a homolytic ground-state
dissociation path.
These observations are in line with the difference between experimental \cite{johnson1999nist} 
gas-phase ionization potentials of H$_2$S and H being (10.5-13.6) eV = -3.1 eV
\footnote{Since $\mathrm{E}(\ce{H2S+ + H})$ - $\mathrm{E}(\ce{H2S + H+})$ = $\mathrm{[ E(H_2S^+)}$ - $\mathrm{E(H_2S) ]}$ + $\mathrm{[ E(H)}$ - $\mathrm{E(H^+) ]}$ = $\mathrm{IP(H_2S)}$ - $\mathrm{IP(H)}$, the inequality $\mathrm{IP(H_2S)}$ - $\mathrm{IP(H)<0}$ implies $\mathrm{E(H_2S^+ + H)}$ $<$ $\mathrm{E(H_2S + H^+)}$.}.

EF inaccurately approximates the electronic structure of the H$_2$S$^{+}$ moiety, 
leading to discrepancies between computed and exact values of $q_S$ (panel $e$). 
QSE improves agreement of $q_S$ with FCI, but quantitatively significant differences 
remain ($\sim 0.1 \, \mathrm{a.u.}$, panel $f$).
The qualitative agreement between computed and exact charges is primarily due to methodological approximations, with decoherence on quantum hardware introducing 
additional deviations in the amount of $\sim 0.01 \, \mathrm{a.u.}$ on average.

\subsection{Excited-state simulations}

Dipole DSFs are shown in Figure~\ref{figure:FIG4}.
While positions and strengths of dominant peaks are in qualitative agreement 
between exact and simulated results, noisy simulations show uncertainties on 
excitation energies, which translate into broadening and overlapping of peaks.
Notwithstanding such limited precision, simulations consistently indicate the 
absorption of ultraviolet (UV) radiation by \sulfonium. 
For all geometries, pronounced peaks are present at $\hbar \omega \simeq 20$ eV 
(0.73 \hartree or 0.9 nm), in the high-energy end of UV.
At $R=0.757$, $1.357$ $\angstrom$ the spectrum is supported above
13 eV (0.48 \hartree or 95 nm). 
As $R$ further increases, structures appear at lower energies,
specifically $\hbar \omega \simeq 4$-$10$ eV (0.15-0.37 \hartree).

To interpret the vertical electronic excitations highlighted by dipole DSFs, 
in  Figure~\ref{figure:FIG5} we show low-lying singlet and triplet excited 
states \revision{(respectively  $\Szero$, $\Sone \dots \mathrm{S}_4$ and $\Tone \dots \mathrm{T}_4$ 
in ascending order of energy at large bond-length)}. 
Simulations on classical and quantum devices predict qualitatively correct curves,
with $\Tone$, $\Szero$ degenerate for large $R$ and $\Ttwo$, $\Sone$ more 
than 100 m\hartree above $\Szero$ across dissociation,
albeit with statistical uncertainties of $\sim 5$ m\hartree for noisy simulations.

Vertical singlet-singlet and singlet-triplet gaps computed from Figure~\ref{figure:FIG5} 
are listed in Table~\ref{table:gaps}.
Noiseless simulations tend to overestimate gaps by 10 and 20 m\hartree respectively. 
Noisy classical and hardware simulations are in line with such trends,
but feature statistical uncertainties of 5 m\hartree respectively.
Note also that FCI predicts $\Sone$, $\Stwo$ and $\Tone$, $\Ttwo$ to be degenerate, but these degeneracies are lifted in QSE, even with the \library{statevector} simulation, because the excitation operators are not symmetry-adapted.

\begin{table}[b!]
\begin{tabular}{c|cccc}
\hline\hline
gap         & FCI & \library{s.v.} & \library{qasm} & \device{kolkata} \\
\hline
$\Sone$-$\Szero$ & 0.484 & 0.493 & \error{0.494}{5} & \error{0.492}{5} \\
$\Stwo$-$\Szero$ & 0.484 & 0.501 & \error{0.502}{6} & \error{0.499}{6} \\
$\Tone$-$\Szero$ & 0.405 & 0.417 & \error{0.417}{5} & \error{0.416}{5} \\
$\Ttwo$-$\Szero$ & 0.405 & 0.428 & \error{0.429}{5} & \error{0.426}{6} \\
\hline\hline
\end{tabular}
\caption{\captiontitle{Singlet-singlet and singlet-triplet gaps} 
Vertical singlet-singlet and singlet-triplet gaps from FCI,
and from QSE with singles and doubles on top of the
EF wavefunction using \library{statevector} (abbreviated \library{s.v.}),
\library{qasm}, and \device{kolkata}.
Gaps are computed at the equilibrium geometry $R=1.357 \angstrom$ and listed in Hartree units.
}
\label{table:gaps}
\end{table}

Figure~\ref{figure:FIG5} also shows partial atomic charges on the departing 
H as a function of $R$. Charges exhibit discontinuous behavior around equilibrium 
geometry.
While FCI and noiseless curves are in agreement with each other,
noisy and hardware simulations significantly deviate from FCI and noiseless values,
indicating the sensitivity of excited-state partial atomic charges to \revision{finite measurement error} and decoherence.
Nevertheless, for large $R$, all simulations agree that $\Sone$, $\Tone$ and $\Ttwo$ 
lead to homolytic dissociation, whereas $\Stwo$ leads to heterolytic dissociation.

Comparison between Figures~\ref{figure:FIG4} and \ref{figure:FIG5} indicates
that absorption of UV light at the equilibrium geometry ($1.357$ $\angstrom$) causes transitions to low-lying singlet states $\Sone$ and $\Stwo$ (indeed, the lowest-energy 
peaks of $S(\omega)$ are located at $\hbar\omega = \Delta E_{\Sone,\Szero}$ and
$\Delta E_{\Stwo,\Szero}$), in turn suggesting coexistence of both homolytic and heterolytic 
pathways in the gas-phase dissociation of \sulfonium, described within an active space.

\section{Conclusions}
\label{sec:conclusions}

In this work, we took a step towards delivering physically relevant simulations
on near-term quantum devices. By integrating the EF technique for qubit reduction 
with quantum diagonalization algorithms exemplified by QSE, we computed ground-
and excited-state wavefunctions and properties of \sulfonium.
Combining these algorithmic developments with a sequence of state-of-the-art 
error mitigation techniques, we experimentally realized the proposed algorithmic
workflow on a superconducting quantum computer.
Note that this is not a direct extension of previous work,
but a careful combination and generalization of independent algorithms.

By computing dipole spectral functions and excited-state energies and partial atomic 
charges, we were able to elucidate the mechanism of dissociation of \sulfonium upon absorption of UV light.

Our study is among the earliest simulations of excited-state molecular spectra
on a quantum processor 
\cite{colless2018computation,gao2021applications,huang2022variational,khan2022chemically}.
Comparison against exact diagonalization indicates that the proposed methodology
is capable to deliver accurate results, at least for the active space sizes 
currently accessible.
Notwithstanding this encouraging result, a number of challenges need to 
be addressed to provide chemically meaningful results, particularly in connection 
with the photo-acid generating properties of \phthrees. This goal can be achieved 
by integrating additional functionalities in the algorithmic workflow considered here. 

This study simulates electrons in an active space derived from
a minimal basis. While simulations of this kind provide benchmarks and occasions
to illustrate algorithmic workflows, useful quantum simulations require accounting
for static and dynamical electron correlation in realistic basis sets; on near-term
devices, this can be achieved using N-electron valence perturbation theory or 
otherwise approximate techniques \cite{tammaro2022n,takeshita2020increasing,boyn2021quantum}.
Industrially relevant photodissociation processes require describing realistic 
functional groups such as phenyl, which on near-term devices can be done integrating 
the algorithms presented here with quasi-complete active space \cite{nakano2000quasi,nakano2001second} or fragmentation techniques
\cite{kawashima2021optimizing,ma2020quantum}.
Simulations are carried out in the gas phase, whereas photo-dissociation reactions 
may occur in a solvent. Solvation effects can be accounted for using implicit or 
explicit solvation models \cite{cheng2020application,castaldo2021quantum}.
Research into these algorithmic extensions and improvements is underway.

Encouragingly, the algorithmic workflow considered in this work appears useful, in 
conjunction with near-term quantum architectures and in combination with other
algorithms, and serves to demonstrate the usefulness of hybrid quantum-classical
simulation techniques in the continuing search for physically relevant simulations.

\section*{Acknowledgments}

We thank Agata M. Bra\'nczyk and Iskandar Sitdikov for generous help and guidance in understanding 
and using a computational package implementing the entanglement forging method.
We acknowledge Riddhi Gupta for valuable interactions regarding the \library{runtime} library.
We thank Joseph Latone for access to the Clifford cluster at IBM Almaden Research Center, where classical computations were carried out.
We thank Andrew Eddins, Hajime Nakamura, Yukio Kawashima, Paul Nation, Zhendong Li, and Barbara Jones for helpful feedback about the manuscript.

\appendix

\section{Additional methodological details}
\label{sec:app_a}

For brevity, the Einstein summation convention is used in most of the equations, while vector expansions, linear combinations of tensor products, and summation over nuclei are explicitly described with the summation symbol.

\subsection{Classical pre-processing}

For each value of the hydrogen-sulfur bondlength studied in this work, 
we performed a constrained geometry optimization 
using density functional theory (DFT) with B3LYP-D3 functional and a def2-qzvpp basis set.

For each geometry, we carried out a restricted closed-shell
Hartree-Fock (RHF or SCF) calculation with the quantum chemistry
software \library{PySCF} at STO-6G level, yielding a set of orthonormal molecular orbitals (MOs) 
$|\psi_p \rangle = \sum_\alpha C_{\alpha p} | \chi_\alpha \rangle$, 
where $| \chi_\alpha \rangle$ are atomic orbitals AOs,
and a second-quantization Born-Oppenheimer Hamiltonian of the form
\begin{equation}
\label{eq:bo_hamiltonian}
\hat{H} = E_0 
+ h_{pr} \crt{p\sigma} \dst{r\sigma} 
+ \frac{(pr|qs)}{2} \crt{p\sigma} \crt{q\tau} \dst{s\tau} \dst{r\sigma} 
\;,
\end{equation}
where indices $prqs=1\dots m$ label MOs, $\sigma,\tau = \uparrow,\downarrow$ label spin 
polarizations.
The nucleus-nucleus Coulomb interaction is given by
\begin{equation}
E_0 = \sum_{\alpha < \beta }^{N_{nuc}} \frac{Z_\alpha Z_\beta}{\| \vett{R}_\alpha - \vett{R}_\beta \|} \quad,
\end{equation}
where $\vett{R}_\alpha$ and $Z_\alpha$ are the position and atomic number of nucleus $\alpha$. 
The coefficients 
\begin{equation}
\begin{split}
h_{pr} &= \int d {\bf{r}} \, \psi^*_p({\bf{r}}) 
\, \left[ - \frac{1}{2} \, \frac{\partial^2}{\partial \vett{r}^2}  - \sum_{\alpha=1}^{N_{nuc}} \frac{Z_\alpha }{ \| \vett{r} - \vett{R}_\alpha \| } \right] \, \psi_r({\bf{r}}) \\
(pr|qs) &= 
\int d {\bf{r}}_1 \int d {\bf{r}}_2 \,  \frac{ \psi^*_p({\bf{r}}_1) \psi_r({\bf{r}}_1) \, \psi^*_q({\bf{r}}_2) \psi_s({\bf{r}}_2) }{ r_{12} }
\end{split}
\end{equation}
specify the one-electron part of the Hamiltonian and the electron-electron Coulomb 
interaction respectively. 
Hartree units are used throughout, the numbers of spin-up and spin-down electrons and 
nuclei are $N_\uparrow$, $N_\downarrow$, and $N_{nuc}$ respectively.

The Born-Oppenheimer Hamiltonian is restricted to an active space spanned by the 6 
higher-energy MOs, corresponding to linear combinations of S[3p] and H[1s] orbitals.
To this end, we froze the 7 lower-energy MOs with a standard frozen-core procedure,
\begin{equation}
\begin{split}
E_0    &\to E_0    + 2 h_{ii} + 2 (ii|jj) - (ij|ji) \quad, \\
h_{pr} &\to h_{pr} + (pr|ii) - (ir|pi) \quad , \\
\end{split}
\end{equation}
where indices $ij$, $pr$ label core and non-core orbitals respectively.

While larger active spaces are necessary for applications of practical interest, we elected to focus on a 6-orbital active space because 
S is valence-isoelectronic with O, and S is not playing 
a hypervalent role as, for example, in \ce{SO4^{2-}}.
\revision{In Table \ref{table:gs} we show the binding energy, vertical singlet-singlet, and vertical singlet-triplet gaps of \sulfonium using CASCI(6e,6o) and CASCI(8e,7o).
As seen, results are quantitatively but not qualitatively affected by the extension of the active space. 
Similarly, both CASCI calculations predict homolytic dissociation along the ground-state potential energy curve.
}

\begin{table}[b!]
\begin{tabular}{c|cccc}
\hline\hline
quantity & CASCI(6e,6o) & CASCI(8e,7o) \\
\hline
$\Sone$-$\Szero$ & 0.484 & 0.474 \\
$\Stwo$-$\Szero$ & 0.484 & 0.474 \\
$\Tone$-$\Szero$ & 0.405 & 0.403 \\
$\Ttwo$-$\Szero$ & 0.405 & 0.403 \\
BE & 0.164 & 0.157 \\
\hline\hline
\end{tabular}
\caption{\captiontitle{Active space comparison} 
Binding energies (BE) and vertical singlet-singlet, singlet-triplet gaps from 
CASCI(6e,6o) and CASCI(8e,7o). Gaps are computed at the equilibrium geometry $R=1.357 \angstrom$ and listed in Hartree units.
}
\label{table:gs}
\end{table}

\revision{Another limitation of the present study is the use of a minimal STO-6G basis set. In Table \ref{table:ccsd} we show the binding energy, vertical singlet-singlet, and vertical singlet-triplet gaps of \sulfonium using classical coupled-cluster with singles and doubles (CCSD) and equation-of-motion CCSD (EOM-CCSD) for ground- and excited-state calculations respectively, in a minimal STO-6G and a correlation-consistent cc-pVTZ basis. Gaps and binding energies are significantly affected by dynamical correlation, but qualitative features (ordering and degeneracy of  the excited states) are preserved.}

\begin{table}[b!]
\begin{tabular}{c|cccc}
\hline\hline
quantity & STO-6G & cc-pVTZ \\
\hline
$\Sone$-$\Szero$ & 0.475 & 0.359 \\
$\Stwo$-$\Szero$ & 0.475 & 0.359 \\
$\Tone$-$\Szero$ & 0.403 & 0.302 \\
$\Ttwo$-$\Szero$ & 0.403 & 0.302 \\
BE & 0.156 & 0.178 \\
\hline\hline
\end{tabular}
\caption{\captiontitle{Basis set comparison} 
Binding energies (BE) and vertical singlet-singlet, singlet-triplet gaps from CCSD and EOM-CCSD respectively, in a STO-6G and a cc-pVTZ basis set, with 5 frozen orbitals. Gaps are computed at the equilibrium geometry $R=1.357 \angstrom$ and listed in Hartree units.
}
\label{table:ccsd}
\end{table}

\subsection{Representation of operators within EF}
\label{app:ef_oper}

In this Section of the Appendix, we discuss how to represent second-quantization operators as qubit operators, and how to evaluate their expectation values within EF.

\subsubsection{The Hamiltonian}

In this work, we chose to partition molecular spin-orbitals into 
spin-up and spin-down.
A Cholesky decomposition \cite{Beebe:1977:683,Aquilante:2007:194106,
motta2018ab,motta2019efficient} of the electron-repulsion integral, 
$(pr|qs) = L^\gamma_{pr} L^\gamma_{qs}$,
and a reordering of the creation and destruction operators,
\begin{equation}
\crt{p\sigma} \crt{q\tau} \dst{s\tau} \dst{r\sigma} 
= 
\crt{p\sigma} \dst{r\sigma} \crt{q\tau} \dst{s\tau}  
-
\delta_{qr} \delta_{\sigma\tau}
\crt{p\sigma} \dst{s\tau} 
\end{equation}
are used to represent the Hamiltonian Eq.~\eqref{eq:bo_hamiltonian} as
\begin{equation}
\oper{H} 
= \left[ h_{pr} - \frac{(pq|qr)}{2} \right] \crt{p\sigma} \dst{r\sigma} 
+ \frac{L^\gamma_{pr} L^\gamma_{qs}}{2} 
\crt{p\sigma} \dst{r\sigma} \crt{q\tau} \dst{s\tau} 
\quad,
\end{equation}
and to separate operators acting on spin-up and spin-down molecular spin-orbitals as
\begin{equation}
\begin{split}
\oper{H} 
&= 
\oper{A}_\uparrow + \oper{A}_\downarrow + \oper{L}^{\gamma}_\uparrow
\oper{L}^{\gamma}_\downarrow \;, \\
\oper{A}_\sigma 
&= 
\left[ h_{pr} - \frac{(pq|qr)}{2} \right] \crt{p\sigma} \dst{r\sigma}
+ \frac{(pr|qs)}{2} \crt{p\sigma} \dst{r\sigma} \crt{q\sigma}  \dst{s\sigma} \;, \\
\oper{L}^\gamma_\sigma &= L^\gamma_{pr} \crt{p\sigma} \dst{r\sigma} \;. \\
\end{split}
\end{equation}
In Jordan-Wigner representation,
\begin{equation}
\label{eq:ef_ham}
\oper{H} = \oper{A} \otimes \mathbbm{1} 
+ 
\mathbbm{1} \otimes \oper{A} + \oper{L}^{\gamma} \otimes \oper{L}^{\gamma} 
\end{equation}
where $\oper{A}$ and $\oper{L}^{\gamma}$ are the qubit representations of 
$\oper{A}_\uparrow$/$\oper{A}_\downarrow$ and $\hat{L}^{\gamma}_\uparrow$/
$\hat{L}^{\gamma}_\downarrow$ restricted to the first/last $m$ qubits respectively.

\subsubsection{Spin-summed one-body operators}

A simpler formula holds for generic one-body operators,
\begin{equation}
\oper{X} 
= x_{pr} \crt{p\sigma} \dst{r\sigma} 
= \oper{X}_\uparrow + \oper{X}_\downarrow
\;.
\end{equation}
In Jordan-Wigner representation it leaves with
\begin{equation}
\label{eq:forging_aux_1}
\oper{X} = \oper{B} \otimes \mathbbm{1} + 
\mathbbm{1} \otimes \oper{B}
\;,
\end{equation}
where $\oper{B}$ is the qubit representation of 
$\oper{X}_\uparrow$/$\oper{X}_\downarrow$ restricted to the first/last 
$m$ qubits.
Equation~\eqref{eq:forging_aux_1} holds for particle number
$x_{pr} = \delta_{pr}$, the spin-summed one-body density
matrix, $x_{pr} = \delta_{p p_0} \delta_{r r_0}$ for element
$(p_0,r_0)$, and the charge-gauge electronic dipole operator,
\begin{equation}
\begin{split}
{\bf{d}}_{pr} &= 
\int d {\bf{r}} \, \psi^*_p({\bf{r}}) ({\bf{r}}-{\bf{r}}_0) \psi_r({\bf{r}}) \;, \\
{\bf{r}}_0 &= \frac{\sum_\alpha Z_\alpha {\bf{R}}_\alpha}{\sum_\alpha Z_\alpha} \;.
\end{split}
\end{equation}

\subsubsection{Total spin}

The total spin operator is
\begin{equation}
\oper{S}^2 = \oper{S}_- \oper{S}_+ + \oper{S}_z (\oper{S}_z+1)
\end{equation}
where $\oper{S}_- = \crt{p\downarrow} \dst{p\uparrow}$,
$\oper{S}_+ = \oper{S}^\dagger_-$,
and $\oper{S}_z = \crt{p\uparrow} \dst{p\uparrow} - \crt{p\downarrow} \dst{p\downarrow}$.
For a closed-shell wavefunction, $\oper{S}_z (\oper{S}_z+1) =0$, leaving 
\begin{equation}
\oper{S}^2 = 
\crt{p\downarrow} \dst{p\uparrow} \crt{q\uparrow} \dst{q\downarrow} 
=
\hat{N}_\downarrow - 
\crt{p\uparrow} \dst{q\uparrow} \crt{p\downarrow} \dst{q\downarrow} 
\end{equation}
In Jordan-Wigner representation,
\begin{equation}
\oper{S}^2 = \mathbbm{I} \otimes \oper{C} + 
\oper{E}_{pq} \otimes \oper{E}_{pq}
\;,
\end{equation}
where $\oper{C}$ is the qubit representation of 
$\hat{N}_\downarrow$ restricted to the last $m$ qubits,
and $\oper{E}_{pq}$ is the qubit representation of $\crt{p\uparrow} \dst{q\uparrow}$/$\crt{p\downarrow} \dst{q\downarrow}$ 
restricted to the first/last $m$ qubits.

$ $

\subsubsection{QSE operators}

In this work, we chose the QSE excitation operators to be single- and
double-electron excitations, respectively
\begin{equation}
\label{eq:forging_singles}
\oper{E}_{ai,\sigma} = \crt{a \sigma} \dst{i \sigma} \;,\;
\oper{E}_{aibj,\sigma\tau} = \crt{a \sigma} \crt{b \tau} \dst{j \tau} \dst{i \sigma}
\;.
\end{equation}
In Jordan-Wigner representation,
\begin{equation}
\label{eq:forging_doubles}
\oper{E}_{ai,\uparrow} = \oper{E}_{ai} \otimes \mathbbm{I} 
\;,\;
\oper{E}_{ai,\downarrow} = \mathbbm{I} \otimes \oper{E}_{ai} 
\end{equation}
for singles and
\begin{equation}
\begin{split}
\oper{E}_{aibj,\uparrow\uparrow} &= \oper{E}_{aibj} \otimes \mathbbm{I} 
\;,\; \\
\oper{E}_{aibj,\downarrow\downarrow} &= \mathbbm{I} \otimes \oper{E}_{aibj}  
\;,\; \\
\oper{E}_{aibj,\uparrow\downarrow} &= \oper{E}_{ai} \otimes \oper{E}_{bj} 
\end{split}
\end{equation}
for doubles. In Eq.~\eqref{eq:forging_singles} and \eqref{eq:forging_doubles},
$\oper{E}_{ai}$/$\oper{E}_{aibj}$ is the qubit representation of  
$\crt{a \uparrow} \dst{i \uparrow}$/
$\crt{a \uparrow} \crt{b \uparrow} \dst{j \uparrow} \dst{i \uparrow}$
restricted to the first $m$ qubits.

Given two or more operators $\oper{X},\oper{Y}$ of the form 
$\oper{X} = \sum_\mu \oper{A}_\mu \otimes \oper{B}_\mu$ and 
$\oper{Y} = \sum_\nu \oper{C}_\nu \otimes \oper{D}_\nu$,
i.e. compatible with Eq.~\eqref{eq:ef_1},
their product can be written as
\begin{equation}
\label{eq:forging_product}
\oper{X} \oper{Y} = \sum_{\mu\nu} 
\oper{A}_\mu \oper{C}_\nu 
\otimes 
\oper{B}_\mu \oper{D}_\nu 
\;,
\end{equation}
which is also compatible with Eq.~\eqref{eq:ef_1}. 
The construction in Eq.~\eqref{eq:forging_product} is used 
to represent QSE operators as linear combinations of tensor
products. 

\revision{
\subsection{Determinant composition of  the EF Ansatz}

Any wavefunction of $(N_\alpha,N_\beta)$ electrons in $m$ spatial orbitals  can be written as a linear combination of electron configurations
\begin{equation}
| \Psi \rangle = \sum_{ij} \psi_{ij} | {\bf{x}}_i {\bf{y}}_j \rangle 
\;,\;
| {\bf{x}} {\bf{y}} \rangle = \prod_{p=0}^{m-1} \left[ \crt{p\uparrow} \right]^{x_p} \left[ \crt{p\downarrow} \right]^{y_p} | \emptyset \rangle \;,
\end{equation}
which are Slater determinants, mapped onto bitstrings by conventional fermion-to-qubit mappings. Here, we elected to highlight the spin-up and spin-down parts of the configuration (respectively ${\bf{x}}$ and ${\bf{y}}$), but other representations are possible, e.g. a partition based on groups of spatial orbitals.
The EF Ansatz is formally derived from a singular value decomposition $\psi_{ij} = \sum_\mu \sigma_\mu U_{i\mu} V_{j\mu}$, leading to
\begin{equation}
\label{eq:ef_from_svd}
| \Psi \rangle = \sum_{\mu} \lambda_\mu \left[ \sum_i U_{i\mu} | {\bf{x}}_i \rangle \right] \left[ \sum_j V_{j\mu} | {\bf{y}}_j \rangle \right]
=
\sum_{\mu} \lambda_\mu |  u_\mu  \rangle | v_\mu \rangle
\;,
\end{equation}
Thus, the EF Ansatz can reproduce any fermionic wavefunction, provided that $(i)$ all the singular values $\lambda_\mu$ are retained in the representation Eq.~\eqref{eq:ef_from_svd} and $(ii)$ quantum circuits $\hat{U}$ and $\hat{V}$ such that $|  u_\mu  \rangle = \hat{U} |  {\bf{x}}_\mu \rangle$, $|  v_\mu  \rangle = \hat{V} |  {\bf{y}}_\mu \rangle$ are available.

In practice, however, the non-zero singular values may be up to $\min \{ \binom{m}{N_\alpha}, \binom{m}{N_\beta}  \}$, which increases combinatorially with active space size, requiring a truncation. Furthermore, while $\hat{U}$ and $\hat{V}$ can be represented as a product of hop-gates with all-to-all connectivity \cite{eddins2021doubling}, implementing such circuits on near-term devices is technically challenging, requiring to use a heuristic Ansatz.

Determining the expressive power of the EF Ansatz in presence of a singular-value truncation and of a heuristic Ansatz for the quantum circuits $\hat{U}$ and $\hat{V}$ is an open problem. However, an
insightful limiting case can be immediately identified: when the circuits
\begin{equation}
\hat{U} = \hat{V} = e^{\hat{K}} \;,\;
\hat{K} = \sum_{pq,\sigma} K_{pq} \crt{p\sigma} \dst{q\sigma}
\;,\;
\hat{K}^\dagger = - \hat{K}
\end{equation}
are equal to the exponential of an anti-Hermitian one-body operator,
the EF Ansatz reduces to the familiar multi-configuration self-consistent field (MCSCF) quantum chemistry method. In such limiting
case, the Ansatz bitstrings correspond to a set of electronic configurations (MC), and the subsequent circuits perform an orbital optimization within the active space (SCF).
In the present work, we focused on approximations to the ground-state  wavefunction for which ${\bf{x}}_\mu = {\bf{y}}_\mu$. This choice corresponds to closed-shell Slater determinants, and cannot reproduce open-shell singlet or triplet  linear combinations of Slater determinants. In this work, we prepared excited states with such character using QSE (i.e., applying suitable linear combinations of single and double excitations), but for the purpose of improving the accuracy of EF and allowing state-specific excited-state calculations, one must allow ${\bf{x}}_\mu$ and ${\bf{y}}_\mu$ to differ.
}

\section{Details of simulations}
\label{sec:app_b}

\subsection{Ground-state EF calculations}
\label{app:ef_details}

The expectation value of the Hamiltonian is written introducing Eq.~\eqref{eq:ef_ham} 
in Eq.~\eqref{eq:ef_1}.
The resulting expectation value is minimized, as a function of all free parameters 
(hop-gate and orbital-optimization angles and coefficients $\lambda_k$), using an
in-house code \cite{entanglement-forging} interfaced with the classical optimization 
method L-BFGS-B \cite{zhu1997algorithm,morales2011remark} and the \library{statevector} 
simulator of \library{Qiskit}.

The variational optimization of coefficients $\lambda_k$ is performed as follows:
it is observed that the energy is a second-degree polynomial in the variables $\lambda$,
\begin{equation}
E(\parm,\lambda) = \langle \Psi_{\parm} | \oper{H} | \Psi_{\parm} \rangle
= \sum_{kl} \lambda_k \lambda_l h_{kl}(\parm)
\;,
\end{equation}
where the Schmidt matrix
\begin{equation}
\label{eq:schmidt_mat}
\begin{split}
h_{kl}(\parm) &= \langle e_k(\parm) | \oper{H} | e_l(\parm) \rangle 
\;,\; \\
| e_k(\parm) \rangle &= \oper{U}(\parm) | \bs_k \rangle \otimes \oper{U}(\parm) | \bs_k \rangle
\end{split}
\end{equation}
is introduced. Therefore, for a fixed parameter configuration $\parm$, the energy is 
minimized when the coefficients $\lambda$ solve the following Lagrange equations,
\begin{equation}
\frac{\partial L}{\partial \lambda_m} = 0
\;,\;
L =  E(\parm,\lambda) - \varepsilon \sum_k \lambda_k^2
\;,
\end{equation}
where a constraint is introduced to ensure normalization of the EF wavefunction. 
The solution of the Lagrange equations is simply
$h_{ml}(\parm) \lambda_l = \varepsilon \lambda_m$,
and the energy is minimized when $\varepsilon$ is the lowest eigenvalue of 
$h_{ml}(\parm)$.

\subsubsection{Ansatz design}

The Ansatz in Figure~\ref{figure:FIG2} was chosen to attain a balance 
between chemical realism and adequacy for contemporary quantum hardware.
The computational basis states $\bs_k$ are chosen to highlight entanglement between
frontier molecular orbitals, and to ensure that preparation unitaries require linear
qubit connectivity only, as per Figure~\ref{figure:FIG2}c.
Similar considerations are made for the subsequent product of hop-gates. The hop-gate
portion of the circuit consists of two blocks of gates, each acting on a sub-group of
3 qubits.
Such a condition ensures that one of the reference states is preserved by the action 
of the EF circuit, $\oper{U}(\parm) | \bs_0 \rangle = | \bs_0 \rangle$, 
so that $h_{00}(\parm)$ in Eq.~\eqref{eq:schmidt_mat} takes the form
\begin{equation}
h_{00}(\parm) = \Big[ \langle \bs_0 | \otimes \langle \bs_0 | \Big] \oper{H} \Big[ | \bs_0 \rangle \otimes | \bs_0 \rangle \Big]
\end{equation}
and can thus be computed classically, thereby reducing the effect of decoherence on EF simulations \cite{eddins2021doubling}.

\subsubsection{HOMO-LUMO orbital optimization}

In this study, MOs were not used as active-space basis functions. Instead, we carried 
out an orbital optimization \cite{mizukami2020orbital,sokolov2020quantum} limited to 
the HOMO-LUMO subspace, i.e. $|\psi^\prime_m \rangle = \sum_l R_{ml}(\varphi) |\psi_l \rangle$
where $R_{ml}(\varphi)$ acts as a SU(2) rotation on the HOMO and LUMO orbitals.
The angle $\varphi$ was optimized along with other variational parameters in a 
preliminary set of ground-state EF calculations (see Section \ref{app:ef_details}), 
then the unitary $R$ was used to transform the integrals in the Born-Oppenheimer 
Hamiltonian with a standard transformation.

\subsubsection{Evaluation of observables}

Once the optimal hop-gate angles and orbital-optimization
angles were computed, quantum state tomography was executed
on the EF circuts in Figure~\ref{figure:FIG2}, for the
purpose of characterizing decoherence effects, enabling 
purification error mitigation, and avoiding repeated measurements.
More specifically, $n=6$ qubits were measured in the $3^n$
eigenbases of X, Y, and Z Pauli operators using the circuit-runner
program from the \library{runtime} library of \library{Qiskit}, 
and the $4^n$ entries of the Bloch vector were computed, along 
with their statistical uncertainties for noisy simulators, by 
standard post-processing.
Post-selection was conducted on the probability distributions
of the Pauli measurements. Readout error mitigation and Clifford 
error mitigation were conducted on the individual entries of the 
Bloch vector, and purification on the resulting Bloch vector.

Expectation values of quantities like the energy were computed 
from the entries of the Bloch vector through standard error 
propagation. For example, given an operator
\begin{equation}
\oper{X} = \sum_i x_i \oper{\sigma}_i
\end{equation}
and a Bloch vector describing the noiseless, noisy, or hardware 
simulation of $| \phi^p_{kl} \rangle$ as
\begin{equation}
| \phi^p_{kl} \rangle \langle \phi^p_{kl} | 
\to 2^{-n} \sum_i a_i(p,k,l) \; \oper{\sigma}_i
\;,
\end{equation}
one has
\begin{equation}
\langle \phi^p_{kl} | \oper{X} | \phi^p_{kl} \rangle
= \sum_i \mu_i x_i
\pm 
\sqrt{ \sum_i v_i^2 x^2_i }
\;,
\end{equation}
where $\mu_i,v_i$ are the mean values and statistical uncertainties 
over Bloch vector entries $a_i(p,k,l)$.
Statistical uncertainties are propagated to observable properties 
(i.e. EF expectation values and derived quantities)
with standard error propagation.

\begin{figure*}[t!]
\centering
\includegraphics[width=0.9\textwidth]{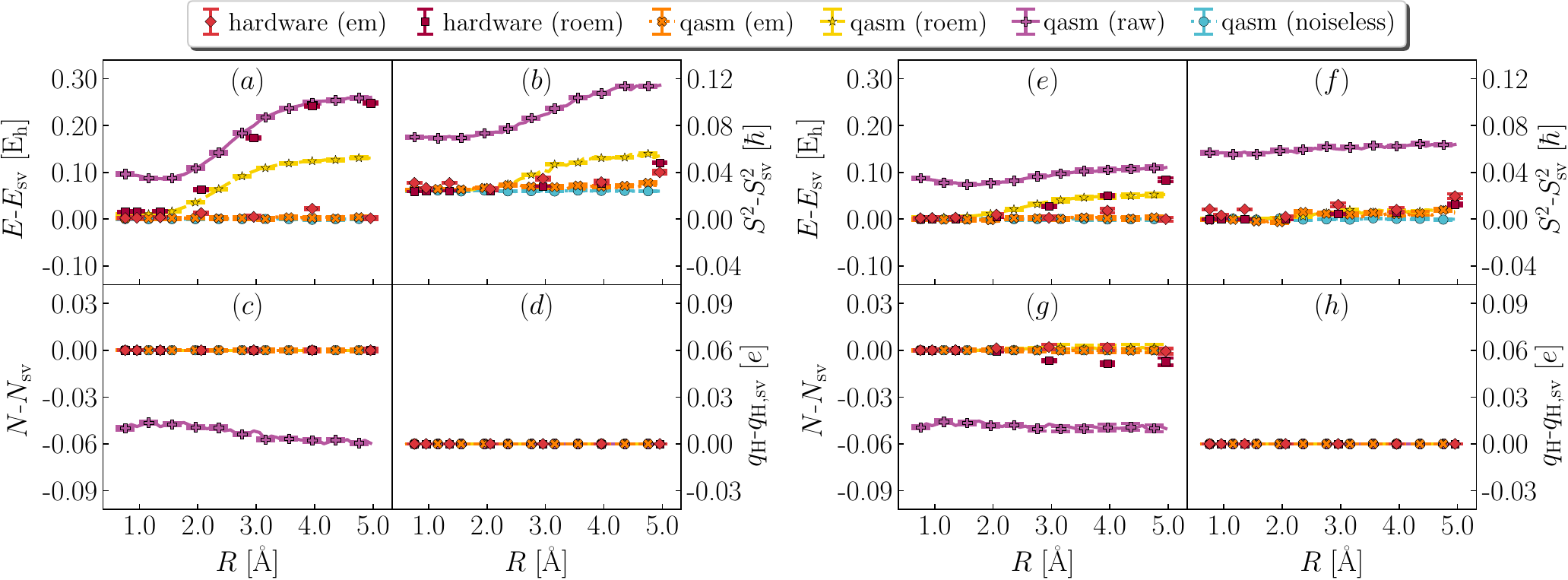}
\caption{ \captiontitle{Effect of error mitigation on ground-state properties} 
Deviation between \library{statevector} and noiseless \library{qasm} (light blue), 
noisy \library{qasm} (purple, yellow and orange for raw, readout error mitigated 
and fully error mitigated data), quantum hardware (\device{kolkata}, dark and 
light red for readout error mitigate and fully error mitigated data) values of 
ground-state energy, total spin, particle number, and atomic charge on the departing 
hydrogen (clockwise), from EF (left) and QSE with singles and doubles on top of the EF 
wavefunction (right). Charges are computed with a Mulliken population analysis based on 
meta-Lowdin atomic orbitals. \important{roem} and \important{em} are abbreviations 
for readout and full error mitigation.
}
\label{figure:FIG6}
\end{figure*}

\subsection{Ground- and excited-state QSE calculations}
\label{app:qse_details}

In the study of Hamiltonian eigenstates with QSE, we
first measured the elements of the metric and total 
spin matrices
\begin{equation}
S_{\mu\nu} = \langle \Psi_{\parm} | \oper{E}_\mu^\dagger \oper{S}^2 \oper{E}_\nu | \Psi_{\parm} \rangle
\;,\;
M_{\mu\nu} = \langle \Psi_{\parm} | \oper{E}_\mu^\dagger \oper{E}_\nu | \Psi_{\parm} \rangle
\;,
\end{equation}
as described in Sections \ref{app:ef_oper} and \ref{app:ef_details}. By solving 
the eigenvalue problem $S_{\mu\nu} f_{\nu t} = M_{\mu\nu} f_{\nu t} \sigma_t$,
QSE matrices are projected on total spin eigenspaces,
\begin{equation}
\begin{split}
H_{\mu\nu} &\to H_{tv} = f_{\mu t} H_{\mu\nu} f_{\nu v}
\;, \\
S_{\mu\nu} &\to S_{tv} = f_{\mu t} S_{\mu\nu} f_{\nu v} =  \delta_{tv} \sigma_t
\;, \\
M_{\mu\nu} &\to M_{tv} = f_{\mu t} M_{\mu\nu} f_{\nu v} = \delta_{tv}
\;, \\
\end{split}
\end{equation}
and the eigenvalue equation $Hg = M g \varepsilon$ is solved within each subspace. 
In summary, QSE energies and spins were defined
introducing $c_{\mu A} = f_{\mu t} \; g_{tA}$ and writing
\begin{equation}
\label{eq:estimators}
\varepsilon_A = 
\frac
{c_{\mu A} H_{\mu\nu} c_{\nu A}}
{c_{\mu A} M_{\mu\nu} c_{\nu A}}
\;,\;
\sigma_A = 
\frac
{c_{\mu A} S_{\mu\nu} c_{\nu A}}
{c_{\mu A} M_{\mu\nu} c_{\nu A}}
\;.
\end{equation}
Statistical uncertainites were assigned to $\varepsilon_A$, $\sigma_A$ starting from
the definition Eq.~\eqref{eq:estimators} as described in the following paragraph.

\subsubsection{Statistical uncertainties}

Determining eigenvalues and eigenvectors of
noisy matrices is a notoriously delicate procedure
\cite{lee2021spectral,blunt2018nonlinear}.
Since eigenvalues of a matrix where elements are normally distributed are not normally distributed, statistical uncertainties are difficult to estimate and are not simply associated with variances of Gaussian distributions. 

In this work, we resorted to a simple numerical
protocol to assign indicative statistical uncertainties to measured quantities:
$(i)$ we solve the eigenvalue equation $S f = M f \sigma$ and $Hg = Mg \varepsilon$
using the mean values of the
matrices $S$ and $M$, without assigning 
statistical uncertainties to solutions $c = fg$.
In this work, we obtained non-singular metric matrices, $\det(M) \gg 0$, so that no eigenvalue truncation was necessary.
$(ii)$ we propagate statistical uncertainties from the matrix elements $H_{\mu\nu}$, $S_{\mu\nu}$,
$M_{\mu\nu}$ to the numerators and denominators
of Eq.~\eqref{eq:estimators}, and to the ratio between these quantities, using standard error propagation.
$(iii)$ an identical procedure was used to assign
statistical uncertainties to particle numbers, RDMs, and derived quantities. Ground- and excited-state RDMs were rescaled so that their trace was statistically compatible with total particle number.

\subsubsection{Evaluation of partial atomic charges}

Upon computing a ground- or excited-state RDM 
(the former with EF, the latter with either EF or QSE),
the RDM was transformed from the active-space to the MO basis
with a simple unitary transformation
$\rho \to R(\varphi) \rho R^\dagger(\varphi)$,
then an extended RDM was generated, by padding the MO-basis
RDM with contributions from frozen MOs,
\begin{equation}
\rho \to \tilde \rho = 
\left(
\begin{array}{c|c}
2 \mathbbm{1} & \mathbf{0} \\
\hline
\mathbf{0} & \rho \\
\end{array}
\right)
\end{equation}
and the extended RDM was transformed to the AO basis,
$\tilde \rho \to C \tilde \rho C^{-1}$. Partial charges
were then computed with a Mulliken population analysis
based on meta-Lowdin atomic orbitals as implemented in
the \library{PySCF} package.
Statistical uncertainties were assigned to partial 
atomic charges by drawing $n=100$ samples of the RDM
$\tilde \rho$ in the AO basis, computing partial atomic
charges for each sample, and averaging results with
standard statistical operations.

\section{Details of hardware simulations}
\label{sec:app_c}

Hardware simulations were carried out on \device{kolkata}.
Jobs submitted on the hardware consisted of 150 circuits 
and $n_s = 100,000$ shots each.
We adapted the circuit-runner program from the \library{runtime} 
library of \library{Qiskit}.
One of the unique options for the circuit-runner 
program is the ability to correct for measurement 
errors (i.e. the \important{roem} technique) 
automatically in the cloud.

\begin{figure}[h!]
\includegraphics[width=\columnwidth]{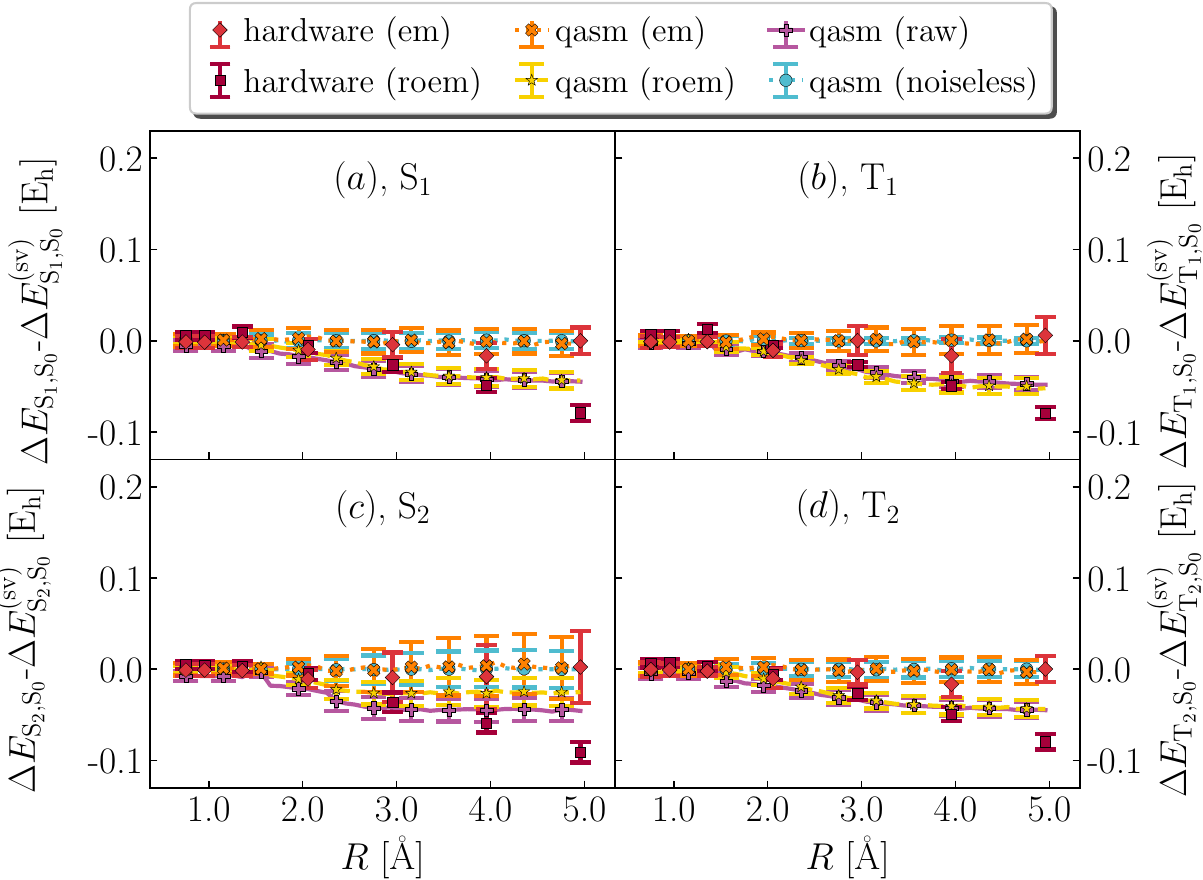}
\caption{ \captiontitle{Effect of error mitigation on excited-state energies} 
Deviation between \library{statevector} and noiseless \library{qasm} (light blue), 
noisy \library{qasm} (purple, yellow and orange for raw, readout error mitigated 
and fully error mitigated data), quantum hardware (\device{kolkata}, dark and light 
red for readout error mitigate and fully error mitigated data)
values of $\mathrm{S_1}$-$\mathrm{S_0}$, $\mathrm{T_1}$-$\mathrm{S_0}$, 
$\mathrm{S_2}$-$\mathrm{S_0}$, $\mathrm{T_2}$-$\mathrm{S_0}$ $(a,b,c,d)$ energy 
differences, from QSE with singles and doubles on top of the EF wavefunction (right).
\important{roem} and \important{em} are abbreviations for readout and full error mitigation.
}
\label{figure:FIG7}
\end{figure}

\subsection{Effect of error mitigation}

In Figures~\ref{figure:FIG6} and \ref{figure:FIG7}, 
we illustrate the impact of various error mitigation techniques on the result of this work. 
In Figure~\ref{figure:FIG6}$a,e$ we focus on ground-state
energies, also shown in Figure~\ref{figure:FIG3} of the main text.
Raw/\important{roem} noisy simulations on \library{qasm} and \important{roem} hardware 
simulations differ by hundreds of m\hartree from noiseless results.
Comparison between \library{qasm} (\important{roem}) and \device{kolkata} (\important{roem}) 
results indicates that noise models underestimate the impact of decoherence on observable 
properties, and comparison between panels $(a)$ and $(e)$ shows that QSE errors are less 
pronounced than EF errors, on average.

In Figure~\ref{figure:FIG6}$b,f$ we focus on total spins, and observe that decoherence 
causes a form of singlet-triplet spin contamination, greatly reduced by error mitigation.
In Figure~\ref{figure:FIG6}$c,g$ we focus on particle number. Owing to post-selection, 
this quantity is essentially noise-free even at the level of \important{roem} data. 
In Figure~\ref{figure:FIG6}$d,h$ we show instead partial charges on the departing H. 
Results indicate that partial charges. While this level of accuracy is satisfactory for 
qualitative applications (e.g. determining the distribution of electric charge along 
dissociation), higher accuracy is needed for quantitative tasks, such as determination 
of electrostatic properties.
Figure \ref{figure:FIG7} shows the effect of error mitigation technique on 
excited-state energies. Decoherence tends to underestimate singlet-singlet and 
singlet-triplet gaps, especially for large $R$.
Error mitigation restores agreement with noiseless results,
but excited-state energies and derived gaps feature large statistical uncertainties, 
as documented in Table \ref{table:gaps} of the main text. The reduction of such 
statistical uncertainties is an important direction of future research.

\bibliographystyle{apsrev4-2}
\bibliography{main}

\begin{thebibliography}{85}%
\makeatletter
\providecommand \@ifxundefined [1]{%
 \@ifx{#1\undefined}
}%
\providecommand \@ifnum [1]{%
 \ifnum #1\expandafter \@firstoftwo
 \else \expandafter \@secondoftwo
 \fi
}%
\providecommand \@ifx [1]{%
 \ifx #1\expandafter \@firstoftwo
 \else \expandafter \@secondoftwo
 \fi
}%
\providecommand \natexlab [1]{#1}%
\providecommand \enquote  [1]{``#1''}%
\providecommand \bibnamefont  [1]{#1}%
\providecommand \bibfnamefont [1]{#1}%
\providecommand \citenamefont [1]{#1}%
\providecommand \href@noop [0]{\@secondoftwo}%
\providecommand \href [0]{\begingroup \@sanitize@url \@href}%
\providecommand \@href[1]{\@@startlink{#1}\@@href}%
\providecommand \@@href[1]{\endgroup#1\@@endlink}%
\providecommand \@sanitize@url [0]{\catcode `\\12\catcode `\$12\catcode
  `\&12\catcode `\#12\catcode `\^12\catcode `\_12\catcode `\%12\relax}%
\providecommand \@@startlink[1]{}%
\providecommand \@@endlink[0]{}%
\providecommand \url  [0]{\begingroup\@sanitize@url \@url }%
\providecommand \@url [1]{\endgroup\@href {#1}{\urlprefix }}%
\providecommand \urlprefix  [0]{URL }%
\providecommand \Eprint [0]{\href }%
\providecommand \doibase [0]{https://doi.org/}%
\providecommand \selectlanguage [0]{\@gobble}%
\providecommand \bibinfo  [0]{\@secondoftwo}%
\providecommand \bibfield  [0]{\@secondoftwo}%
\providecommand \translation [1]{[#1]}%
\providecommand \BibitemOpen [0]{}%
\providecommand \bibitemStop [0]{}%
\providecommand \bibitemNoStop [0]{.\EOS\space}%
\providecommand \EOS [0]{\spacefactor3000\relax}%
\providecommand \BibitemShut  [1]{\csname bibitem#1\endcsname}%
\let\auto@bib@innerbib\@empty
\bibitem [{\citenamefont {Helgaker}\ \emph {et~al.}(2012)\citenamefont
  {Helgaker}, \citenamefont {Coriani}, \citenamefont {J{\o}rgensen},
  \citenamefont {Kristensen}, \citenamefont {Olsen},\ and\ \citenamefont
  {Ruud}}]{helgaker2012recent}%
  \BibitemOpen
  \bibfield  {author} {\bibinfo {author} {\bibfnamefont {T.}~\bibnamefont
  {Helgaker}}, \bibinfo {author} {\bibfnamefont {S.}~\bibnamefont {Coriani}},
  \bibinfo {author} {\bibfnamefont {P.}~\bibnamefont {J{\o}rgensen}}, \bibinfo
  {author} {\bibfnamefont {K.}~\bibnamefont {Kristensen}}, \bibinfo {author}
  {\bibfnamefont {J.}~\bibnamefont {Olsen}},\ and\ \bibinfo {author}
  {\bibfnamefont {K.}~\bibnamefont {Ruud}},\ }\href
  {https://pubs.acs.org/doi/10.1021/cr2002239} {\bibfield  {journal} {\bibinfo
  {journal} {Chem. Rev}\ }\textbf {\bibinfo {volume} {112}},\ \bibinfo {pages}
  {543} (\bibinfo {year} {2012})}\BibitemShut {NoStop}%
\bibitem [{\citenamefont {Friesner}(2005)}]{friesner2005ab}%
  \BibitemOpen
  \bibfield  {author} {\bibinfo {author} {\bibfnamefont {R.~A.}\ \bibnamefont
  {Friesner}},\ }\href {https://www.pnas.org/doi/10.1073/pnas.0408036102}
  {\bibfield  {journal} {\bibinfo  {journal} {Proc. Natl. Acad. Sci}\ }\textbf
  {\bibinfo {volume} {102}},\ \bibinfo {pages} {6648} (\bibinfo {year}
  {2005})}\BibitemShut {NoStop}%
\bibitem [{\citenamefont {Gates}\ \emph {et~al.}(2005)\citenamefont {Gates},
  \citenamefont {Xu}, \citenamefont {Stewart}, \citenamefont {Ryan},
  \citenamefont {Willson},\ and\ \citenamefont {Whitesides}}]{gates2005new}%
  \BibitemOpen
  \bibfield  {author} {\bibinfo {author} {\bibfnamefont {B.~D.}\ \bibnamefont
  {Gates}}, \bibinfo {author} {\bibfnamefont {Q.}~\bibnamefont {Xu}}, \bibinfo
  {author} {\bibfnamefont {M.}~\bibnamefont {Stewart}}, \bibinfo {author}
  {\bibfnamefont {D.}~\bibnamefont {Ryan}}, \bibinfo {author} {\bibfnamefont
  {C.~G.}\ \bibnamefont {Willson}},\ and\ \bibinfo {author} {\bibfnamefont
  {G.~M.}\ \bibnamefont {Whitesides}},\ }\href
  {https://pubs.acs.org/doi/10.1021/cr030076o} {\bibfield  {journal} {\bibinfo
  {journal} {Chem. Rev}\ }\textbf {\bibinfo {volume} {105}},\ \bibinfo {pages}
  {1171} (\bibinfo {year} {2005})}\BibitemShut {NoStop}%
\bibitem [{\citenamefont {Gangnaik}\ \emph {et~al.}(2017)\citenamefont
  {Gangnaik}, \citenamefont {Georgiev},\ and\ \citenamefont
  {Holmes}}]{gangnaik2017new}%
  \BibitemOpen
  \bibfield  {author} {\bibinfo {author} {\bibfnamefont {A.~S.}\ \bibnamefont
  {Gangnaik}}, \bibinfo {author} {\bibfnamefont {Y.~M.}\ \bibnamefont
  {Georgiev}},\ and\ \bibinfo {author} {\bibfnamefont {J.~D.}\ \bibnamefont
  {Holmes}},\ }\href {https://pubs.acs.org/doi/10.1021/acs.chemmater.6b03483}
  {\bibfield  {journal} {\bibinfo  {journal} {Chem. Mat}\ }\textbf {\bibinfo
  {volume} {29}},\ \bibinfo {pages} {1898} (\bibinfo {year}
  {2017})}\BibitemShut {NoStop}%
\bibitem [{\citenamefont {Nalamasu}\ \emph {et~al.}(1990)\citenamefont
  {Nalamasu}, \citenamefont {Cheng}, \citenamefont {Kometani}, \citenamefont
  {Vaidya}, \citenamefont {Reichmanis},\ and\ \citenamefont
  {Thompson}}]{nalamasu1990development}%
  \BibitemOpen
  \bibfield  {author} {\bibinfo {author} {\bibfnamefont {O.}~\bibnamefont
  {Nalamasu}}, \bibinfo {author} {\bibfnamefont {M.}~\bibnamefont {Cheng}},
  \bibinfo {author} {\bibfnamefont {J.~M.}\ \bibnamefont {Kometani}}, \bibinfo
  {author} {\bibfnamefont {S.}~\bibnamefont {Vaidya}}, \bibinfo {author}
  {\bibfnamefont {E.}~\bibnamefont {Reichmanis}},\ and\ \bibinfo {author}
  {\bibfnamefont {L.~F.}\ \bibnamefont {Thompson}},\ }in\ \href@noop {} {\emph
  {\bibinfo {booktitle} {Advances in Resist Technology and Processing VII}}},\
  Vol.\ \bibinfo {volume} {1262}\ (\bibinfo {organization} {SPIE},\ \bibinfo
  {year} {1990})\ pp.\ \bibinfo {pages} {32--48}\BibitemShut {NoStop}%
\bibitem [{\citenamefont {Fallica}\ and\ \citenamefont
  {Ekinci}(2018)}]{fallica2018photoacid}%
  \BibitemOpen
  \bibfield  {author} {\bibinfo {author} {\bibfnamefont {R.}~\bibnamefont
  {Fallica}}\ and\ \bibinfo {author} {\bibfnamefont {Y.}~\bibnamefont
  {Ekinci}},\ }\href
  {https://pubs.rsc.org/en/content/articlelanding/2018/tc/c8tc01446a}
  {\bibfield  {journal} {\bibinfo  {journal} {J. Mat. Chem. C}\ }\textbf
  {\bibinfo {volume} {6}},\ \bibinfo {pages} {7267} (\bibinfo {year}
  {2018})}\BibitemShut {NoStop}%
\bibitem [{\citenamefont {Martin}\ \emph {et~al.}(2018)\citenamefont {Martin},
  \citenamefont {Rapenne}, \citenamefont {Nakashima},\ and\ \citenamefont
  {Kawai}}]{martin2018recent}%
  \BibitemOpen
  \bibfield  {author} {\bibinfo {author} {\bibfnamefont {C.~J.}\ \bibnamefont
  {Martin}}, \bibinfo {author} {\bibfnamefont {G.}~\bibnamefont {Rapenne}},
  \bibinfo {author} {\bibfnamefont {T.}~\bibnamefont {Nakashima}},\ and\
  \bibinfo {author} {\bibfnamefont {T.}~\bibnamefont {Kawai}},\ }\href
  {https://www.sciencedirect.com/science/article/abs/pii/S1389556717300734}
  {\bibfield  {journal} {\bibinfo  {journal} {J. Photochem. Photobiol. C}\
  }\textbf {\bibinfo {volume} {34}},\ \bibinfo {pages} {41} (\bibinfo {year}
  {2018})}\BibitemShut {NoStop}%
\bibitem [{\citenamefont {Sambath}\ \emph {et~al.}(2020)\citenamefont
  {Sambath}, \citenamefont {Wan}, \citenamefont {Wang}, \citenamefont {Chen},\
  and\ \citenamefont {Zhang}}]{sambath2020bodipy}%
  \BibitemOpen
  \bibfield  {author} {\bibinfo {author} {\bibfnamefont {K.}~\bibnamefont
  {Sambath}}, \bibinfo {author} {\bibfnamefont {Z.}~\bibnamefont {Wan}},
  \bibinfo {author} {\bibfnamefont {Q.}~\bibnamefont {Wang}}, \bibinfo {author}
  {\bibfnamefont {H.}~\bibnamefont {Chen}},\ and\ \bibinfo {author}
  {\bibfnamefont {Y.}~\bibnamefont {Zhang}},\ }\href
  {https://pubs.acs.org/doi/10.1021/acs.orglett.0c00118} {\bibfield  {journal}
  {\bibinfo  {journal} {Org. Lett}\ }\textbf {\bibinfo {volume} {22}},\
  \bibinfo {pages} {1208} (\bibinfo {year} {2020})}\BibitemShut {NoStop}%
\bibitem [{\citenamefont {Ohmori}\ \emph {et~al.}(1998)\citenamefont {Ohmori},
  \citenamefont {Nakazono}, \citenamefont {Hata}, \citenamefont {Hoshino},\
  and\ \citenamefont {Tsuda}}]{ohmori1998ab}%
  \BibitemOpen
  \bibfield  {author} {\bibinfo {author} {\bibfnamefont {N.}~\bibnamefont
  {Ohmori}}, \bibinfo {author} {\bibfnamefont {Y.}~\bibnamefont {Nakazono}},
  \bibinfo {author} {\bibfnamefont {M.}~\bibnamefont {Hata}}, \bibinfo {author}
  {\bibfnamefont {T.}~\bibnamefont {Hoshino}},\ and\ \bibinfo {author}
  {\bibfnamefont {M.}~\bibnamefont {Tsuda}},\ }\href
  {https://pubs.acs.org/doi/10.1021/jp9726100} {\bibfield  {journal} {\bibinfo
  {journal} {J. Phys. Chem. B}\ }\textbf {\bibinfo {volume} {102}},\ \bibinfo
  {pages} {927} (\bibinfo {year} {1998})}\BibitemShut {NoStop}%
\bibitem [{\citenamefont {Dektar}\ and\ \citenamefont
  {Hacker}(1988)}]{dektar1988triphenylsulfonium}%
  \BibitemOpen
  \bibfield  {author} {\bibinfo {author} {\bibfnamefont {J.~L.}\ \bibnamefont
  {Dektar}}\ and\ \bibinfo {author} {\bibfnamefont {N.~P.}\ \bibnamefont
  {Hacker}},\ }\href {https://pubs.acs.org/doi/10.1021/jo00243a053} {\bibfield
  {journal} {\bibinfo  {journal} {J. Org. Chem}\ }\textbf {\bibinfo {volume}
  {53}},\ \bibinfo {pages} {1833} (\bibinfo {year} {1988})}\BibitemShut
  {NoStop}%
\bibitem [{\citenamefont {Dektar}\ and\ \citenamefont
  {Hacker}(1990)}]{dektar1990photochemistry}%
  \BibitemOpen
  \bibfield  {author} {\bibinfo {author} {\bibfnamefont {J.~L.}\ \bibnamefont
  {Dektar}}\ and\ \bibinfo {author} {\bibfnamefont {N.~P.}\ \bibnamefont
  {Hacker}},\ }\href {https://pubs.acs.org/doi/10.1021/ja00172a015} {\bibfield
  {journal} {\bibinfo  {journal} {J. Am. Chem. Soc}\ }\textbf {\bibinfo
  {volume} {112}},\ \bibinfo {pages} {6004} (\bibinfo {year}
  {1990})}\BibitemShut {NoStop}%
\bibitem [{\citenamefont {Klikovits}\ \emph {et~al.}(2017)\citenamefont
  {Klikovits}, \citenamefont {Knaack}, \citenamefont {Bomze}, \citenamefont
  {Krossing},\ and\ \citenamefont {Liska}}]{klikovits2017novel}%
  \BibitemOpen
  \bibfield  {author} {\bibinfo {author} {\bibfnamefont {N.}~\bibnamefont
  {Klikovits}}, \bibinfo {author} {\bibfnamefont {P.}~\bibnamefont {Knaack}},
  \bibinfo {author} {\bibfnamefont {D.}~\bibnamefont {Bomze}}, \bibinfo
  {author} {\bibfnamefont {I.}~\bibnamefont {Krossing}},\ and\ \bibinfo
  {author} {\bibfnamefont {R.}~\bibnamefont {Liska}},\ }\href
  {https://pubs.rsc.org/en/content/articlelanding/2017/PY/C7PY00855D}
  {\bibfield  {journal} {\bibinfo  {journal} {Polym. Chem}\ }\textbf {\bibinfo
  {volume} {8}},\ \bibinfo {pages} {4414} (\bibinfo {year} {2017})}\BibitemShut
  {NoStop}%
\bibitem [{\citenamefont {Jin}\ \emph {et~al.}(2014)\citenamefont {Jin},
  \citenamefont {Hong}, \citenamefont {Xie}, \citenamefont {Malval},
  \citenamefont {Spangenberg}, \citenamefont {Soppera}, \citenamefont {Wan},
  \citenamefont {Pu}, \citenamefont {Versace}, \citenamefont {Leclerc} \emph
  {et~al.}}]{jin2014pi}%
  \BibitemOpen
  \bibfield  {author} {\bibinfo {author} {\bibfnamefont {M.}~\bibnamefont
  {Jin}}, \bibinfo {author} {\bibfnamefont {H.}~\bibnamefont {Hong}}, \bibinfo
  {author} {\bibfnamefont {J.}~\bibnamefont {Xie}}, \bibinfo {author}
  {\bibfnamefont {J.-P.}\ \bibnamefont {Malval}}, \bibinfo {author}
  {\bibfnamefont {A.}~\bibnamefont {Spangenberg}}, \bibinfo {author}
  {\bibfnamefont {O.}~\bibnamefont {Soppera}}, \bibinfo {author} {\bibfnamefont
  {D.}~\bibnamefont {Wan}}, \bibinfo {author} {\bibfnamefont {H.}~\bibnamefont
  {Pu}}, \bibinfo {author} {\bibfnamefont {D.-L.}\ \bibnamefont {Versace}},
  \bibinfo {author} {\bibfnamefont {T.}~\bibnamefont {Leclerc}}, \emph
  {et~al.},\ }\href
  {https://pubs.rsc.org/en/content/articlelanding/2014/py/c4py00424h}
  {\bibfield  {journal} {\bibinfo  {journal} {Polym. Chem}\ }\textbf {\bibinfo
  {volume} {5}},\ \bibinfo {pages} {4747} (\bibinfo {year} {2014})}\BibitemShut
  {NoStop}%
\bibitem [{\citenamefont {Zhou}\ \emph {et~al.}(2002)\citenamefont {Zhou},
  \citenamefont {Kuebler}, \citenamefont {Carrig}, \citenamefont {Perry},\ and\
  \citenamefont {Marder}}]{zhou2002efficient}%
  \BibitemOpen
  \bibfield  {author} {\bibinfo {author} {\bibfnamefont {W.}~\bibnamefont
  {Zhou}}, \bibinfo {author} {\bibfnamefont {S.~M.}\ \bibnamefont {Kuebler}},
  \bibinfo {author} {\bibfnamefont {D.}~\bibnamefont {Carrig}}, \bibinfo
  {author} {\bibfnamefont {J.~W.}\ \bibnamefont {Perry}},\ and\ \bibinfo
  {author} {\bibfnamefont {S.~R.}\ \bibnamefont {Marder}},\ }\href
  {https://pubs.acs.org/doi/10.1021/ja011186k} {\bibfield  {journal} {\bibinfo
  {journal} {J. Am. Chem. Soc}\ }\textbf {\bibinfo {volume} {124}},\ \bibinfo
  {pages} {1897} (\bibinfo {year} {2002})}\BibitemShut {NoStop}%
\bibitem [{\citenamefont {Knapczyk}\ and\ \citenamefont
  {McEwen}(1969)}]{knapczyk1969reactions}%
  \BibitemOpen
  \bibfield  {author} {\bibinfo {author} {\bibfnamefont {J.~W.}\ \bibnamefont
  {Knapczyk}}\ and\ \bibinfo {author} {\bibfnamefont {W.~E.}\ \bibnamefont
  {McEwen}},\ }\href {https://pubs.acs.org/doi/pdf/10.1021/ja01029a029}
  {\bibfield  {journal} {\bibinfo  {journal} {J. Am. Chem. Soc}\ }\textbf
  {\bibinfo {volume} {91}},\ \bibinfo {pages} {145} (\bibinfo {year}
  {1969})}\BibitemShut {NoStop}%
\bibitem [{\citenamefont {Crivello}\ and\ \citenamefont
  {Lam}(1979)}]{crivello1979dye}%
  \BibitemOpen
  \bibfield  {author} {\bibinfo {author} {\bibfnamefont {J.}~\bibnamefont
  {Crivello}}\ and\ \bibinfo {author} {\bibfnamefont {J.}~\bibnamefont {Lam}},\
  }\href {https://onlinelibrary.wiley.com/doi/10.1002/pol.1979.170170411}
  {\bibfield  {journal} {\bibinfo  {journal} {J. Polym. Sci., Polym. Chem. Ed}\
  }\textbf {\bibinfo {volume} {17}},\ \bibinfo {pages} {1059} (\bibinfo {year}
  {1979})}\BibitemShut {NoStop}%
\bibitem [{\citenamefont {Pappas}\ \emph {et~al.}(1984)\citenamefont {Pappas},
  \citenamefont {Pappas}, \citenamefont {Gatechair}, \citenamefont {Jilek},\
  and\ \citenamefont {Schnabel}}]{pappas1984photoinitiation}%
  \BibitemOpen
  \bibfield  {author} {\bibinfo {author} {\bibfnamefont {S.~P.}\ \bibnamefont
  {Pappas}}, \bibinfo {author} {\bibfnamefont {B.~C.}\ \bibnamefont {Pappas}},
  \bibinfo {author} {\bibfnamefont {L.~R.}\ \bibnamefont {Gatechair}}, \bibinfo
  {author} {\bibfnamefont {J.~H.}\ \bibnamefont {Jilek}},\ and\ \bibinfo
  {author} {\bibfnamefont {W.}~\bibnamefont {Schnabel}},\ }\href
  {https://www.sciencedirect.com/science/article/abs/pii/0144288084900186}
  {\bibfield  {journal} {\bibinfo  {journal} {Polym. Photochem}\ }\textbf
  {\bibinfo {volume} {5}},\ \bibinfo {pages} {1} (\bibinfo {year}
  {1984})}\BibitemShut {NoStop}%
\bibitem [{\citenamefont {Davidson}\ and\ \citenamefont
  {Goodin}(1982)}]{davidson1982some}%
  \BibitemOpen
  \bibfield  {author} {\bibinfo {author} {\bibfnamefont {R.}~\bibnamefont
  {Davidson}}\ and\ \bibinfo {author} {\bibfnamefont {J.}~\bibnamefont
  {Goodin}},\ }\href
  {https://www.sciencedirect.com/science/article/abs/pii/0014305782900362}
  {\bibfield  {journal} {\bibinfo  {journal} {Eur. Polym. J}\ }\textbf
  {\bibinfo {volume} {18}},\ \bibinfo {pages} {589} (\bibinfo {year}
  {1982})}\BibitemShut {NoStop}%
\bibitem [{\citenamefont {LeBlanc}\ \emph {et~al.}(2015)\citenamefont {LeBlanc}
  \emph {et~al.}}]{LeBlanc_PRX_2015}%
  \BibitemOpen
  \bibfield  {author} {\bibinfo {author} {\bibfnamefont {J.~P.~F.}\
  \bibnamefont {LeBlanc}} \emph {et~al.},\ }\href
  {https://doi.org/10.1103/PhysRevX.5.041041} {\bibfield  {journal} {\bibinfo
  {journal} {Phys. Rev. X}\ }\textbf {\bibinfo {volume} {5}},\ \bibinfo {pages}
  {041041} (\bibinfo {year} {2015})}\BibitemShut {NoStop}%
\bibitem [{\citenamefont {Zheng}\ \emph {et~al.}(2017)\citenamefont {Zheng},
  \citenamefont {Chung}, \citenamefont {Corboz}, \citenamefont {Ehlers},
  \citenamefont {Qin}, \citenamefont {Noack}, \citenamefont {Shi},
  \citenamefont {White}, \citenamefont {Zhang},\ and\ \citenamefont
  {Chan}}]{Zheng_Science_2017}%
  \BibitemOpen
  \bibfield  {author} {\bibinfo {author} {\bibfnamefont {B.-X.}\ \bibnamefont
  {Zheng}}, \bibinfo {author} {\bibfnamefont {C.-M.}\ \bibnamefont {Chung}},
  \bibinfo {author} {\bibfnamefont {P.}~\bibnamefont {Corboz}}, \bibinfo
  {author} {\bibfnamefont {G.}~\bibnamefont {Ehlers}}, \bibinfo {author}
  {\bibfnamefont {M.-P.}\ \bibnamefont {Qin}}, \bibinfo {author} {\bibfnamefont
  {R.~M.}\ \bibnamefont {Noack}}, \bibinfo {author} {\bibfnamefont
  {H.}~\bibnamefont {Shi}}, \bibinfo {author} {\bibfnamefont {S.~R.}\
  \bibnamefont {White}}, \bibinfo {author} {\bibfnamefont {S.}~\bibnamefont
  {Zhang}},\ and\ \bibinfo {author} {\bibfnamefont {G.~K.-L.}\ \bibnamefont
  {Chan}},\ }\href {https://doi.org/10.1126/science.aam7127} {\bibfield
  {journal} {\bibinfo  {journal} {Science}\ }\textbf {\bibinfo {volume}
  {358}},\ \bibinfo {pages} {1155} (\bibinfo {year} {2017})}\BibitemShut
  {NoStop}%
\bibitem [{\citenamefont {Motta}\ \emph {et~al.}(2017)\citenamefont {Motta},
  \citenamefont {Ceperley}, \citenamefont {Chan}, \citenamefont {Gomez},
  \citenamefont {Gull}, \citenamefont {Guo}, \citenamefont {Jim\'enez-Hoyos},
  \citenamefont {Lan}, \citenamefont {Li}, \citenamefont {Ma}, \citenamefont
  {Millis}, \citenamefont {Prokof'ev}, \citenamefont {Ray}, \citenamefont
  {Scuseria}, \citenamefont {Sorella}, \citenamefont {Stoudenmire},
  \citenamefont {Sun}, \citenamefont {Tupitsyn}, \citenamefont {White},
  \citenamefont {Zgid},\ and\ \citenamefont {Zhang}}]{Motta_PRX_2017}%
  \BibitemOpen
  \bibfield  {author} {\bibinfo {author} {\bibfnamefont {M.}~\bibnamefont
  {Motta}}, \bibinfo {author} {\bibfnamefont {D.~M.}\ \bibnamefont {Ceperley}},
  \bibinfo {author} {\bibfnamefont {G.~K.-L.}\ \bibnamefont {Chan}}, \bibinfo
  {author} {\bibfnamefont {J.~A.}\ \bibnamefont {Gomez}}, \bibinfo {author}
  {\bibfnamefont {E.}~\bibnamefont {Gull}}, \bibinfo {author} {\bibfnamefont
  {S.}~\bibnamefont {Guo}}, \bibinfo {author} {\bibfnamefont {C.~A.}\
  \bibnamefont {Jim\'enez-Hoyos}}, \bibinfo {author} {\bibfnamefont {T.~N.}\
  \bibnamefont {Lan}}, \bibinfo {author} {\bibfnamefont {J.}~\bibnamefont
  {Li}}, \bibinfo {author} {\bibfnamefont {F.}~\bibnamefont {Ma}}, \bibinfo
  {author} {\bibfnamefont {A.~J.}\ \bibnamefont {Millis}}, \bibinfo {author}
  {\bibfnamefont {N.~V.}\ \bibnamefont {Prokof'ev}}, \bibinfo {author}
  {\bibfnamefont {U.}~\bibnamefont {Ray}}, \bibinfo {author} {\bibfnamefont
  {G.~E.}\ \bibnamefont {Scuseria}}, \bibinfo {author} {\bibfnamefont
  {S.}~\bibnamefont {Sorella}}, \bibinfo {author} {\bibfnamefont {E.~M.}\
  \bibnamefont {Stoudenmire}}, \bibinfo {author} {\bibfnamefont
  {Q.}~\bibnamefont {Sun}}, \bibinfo {author} {\bibfnamefont {I.~S.}\
  \bibnamefont {Tupitsyn}}, \bibinfo {author} {\bibfnamefont {S.~R.}\
  \bibnamefont {White}}, \bibinfo {author} {\bibfnamefont {D.}~\bibnamefont
  {Zgid}},\ and\ \bibinfo {author} {\bibfnamefont {S.}~\bibnamefont {Zhang}},\
  }\href {https://doi.org/10.1103/PhysRevX.7.031059} {\bibfield  {journal}
  {\bibinfo  {journal} {Phys. Rev. X}\ }\textbf {\bibinfo {volume} {7}},\
  \bibinfo {pages} {031059} (\bibinfo {year} {2017})}\BibitemShut {NoStop}%
\bibitem [{\citenamefont {Williams}\ \emph {et~al.}(2020)\citenamefont
  {Williams}, \citenamefont {Yao}, \citenamefont {Li}, \citenamefont {Chen},
  \citenamefont {Shi}, \citenamefont {Motta}, \citenamefont {Niu},
  \citenamefont {Ray}, \citenamefont {Guo}, \citenamefont {Anderson} \emph
  {et~al.}}]{williams2020direct}%
  \BibitemOpen
  \bibfield  {author} {\bibinfo {author} {\bibfnamefont {K.~T.}\ \bibnamefont
  {Williams}}, \bibinfo {author} {\bibfnamefont {Y.}~\bibnamefont {Yao}},
  \bibinfo {author} {\bibfnamefont {J.}~\bibnamefont {Li}}, \bibinfo {author}
  {\bibfnamefont {L.}~\bibnamefont {Chen}}, \bibinfo {author} {\bibfnamefont
  {H.}~\bibnamefont {Shi}}, \bibinfo {author} {\bibfnamefont {M.}~\bibnamefont
  {Motta}}, \bibinfo {author} {\bibfnamefont {C.}~\bibnamefont {Niu}}, \bibinfo
  {author} {\bibfnamefont {U.}~\bibnamefont {Ray}}, \bibinfo {author}
  {\bibfnamefont {S.}~\bibnamefont {Guo}}, \bibinfo {author} {\bibfnamefont
  {R.~J.}\ \bibnamefont {Anderson}}, \emph {et~al.},\ }\href
  {https://journals.aps.org/prx/abstract/10.1103/PhysRevX.10.011041} {\bibfield
   {journal} {\bibinfo  {journal} {Phys. Rev. X}\ }\textbf {\bibinfo {volume}
  {10}},\ \bibinfo {pages} {011041} (\bibinfo {year} {2020})}\BibitemShut
  {NoStop}%
\bibitem [{\citenamefont {Georgescu}\ \emph {et~al.}(2014)\citenamefont
  {Georgescu}, \citenamefont {Ashhab},\ and\ \citenamefont
  {Nori}}]{georgescu2014quantum}%
  \BibitemOpen
  \bibfield  {author} {\bibinfo {author} {\bibfnamefont {I.~M.}\ \bibnamefont
  {Georgescu}}, \bibinfo {author} {\bibfnamefont {S.}~\bibnamefont {Ashhab}},\
  and\ \bibinfo {author} {\bibfnamefont {F.}~\bibnamefont {Nori}},\ }\href
  {https://journals.aps.org/rmp/abstract/10.1103/RevModPhys.86.153} {\bibfield
  {journal} {\bibinfo  {journal} {Rev. Mod. Phys}\ }\textbf {\bibinfo {volume}
  {86}},\ \bibinfo {pages} {153} (\bibinfo {year} {2014})}\BibitemShut
  {NoStop}%
\bibitem [{\citenamefont {Cao}\ \emph {et~al.}(2019)\citenamefont {Cao},
  \citenamefont {Romero}, \citenamefont {Olson}, \citenamefont {Degroote},
  \citenamefont {Johnson}, \citenamefont {Kieferov{\'a}}, \citenamefont
  {Kivlichan}, \citenamefont {Menke}, \citenamefont {Peropadre}, \citenamefont
  {Sawaya} \emph {et~al.}}]{cao2019quantum}%
  \BibitemOpen
  \bibfield  {author} {\bibinfo {author} {\bibfnamefont {Y.}~\bibnamefont
  {Cao}}, \bibinfo {author} {\bibfnamefont {J.}~\bibnamefont {Romero}},
  \bibinfo {author} {\bibfnamefont {J.~P.}\ \bibnamefont {Olson}}, \bibinfo
  {author} {\bibfnamefont {M.}~\bibnamefont {Degroote}}, \bibinfo {author}
  {\bibfnamefont {P.~D.}\ \bibnamefont {Johnson}}, \bibinfo {author}
  {\bibfnamefont {M.}~\bibnamefont {Kieferov{\'a}}}, \bibinfo {author}
  {\bibfnamefont {I.~D.}\ \bibnamefont {Kivlichan}}, \bibinfo {author}
  {\bibfnamefont {T.}~\bibnamefont {Menke}}, \bibinfo {author} {\bibfnamefont
  {B.}~\bibnamefont {Peropadre}}, \bibinfo {author} {\bibfnamefont {N.~P.}\
  \bibnamefont {Sawaya}}, \emph {et~al.},\ }\href
  {https://pubs.acs.org/doi/10.1021/acs.chemrev.8b00803} {\bibfield  {journal}
  {\bibinfo  {journal} {Chem. Rev}\ }\textbf {\bibinfo {volume} {119}},\
  \bibinfo {pages} {10856} (\bibinfo {year} {2019})}\BibitemShut {NoStop}%
\bibitem [{\citenamefont {Bauer}\ \emph {et~al.}(2020)\citenamefont {Bauer},
  \citenamefont {Bravyi}, \citenamefont {Motta},\ and\ \citenamefont
  {Kin-Lic~Chan}}]{bauer2020quantum}%
  \BibitemOpen
  \bibfield  {author} {\bibinfo {author} {\bibfnamefont {B.}~\bibnamefont
  {Bauer}}, \bibinfo {author} {\bibfnamefont {S.}~\bibnamefont {Bravyi}},
  \bibinfo {author} {\bibfnamefont {M.}~\bibnamefont {Motta}},\ and\ \bibinfo
  {author} {\bibfnamefont {G.}~\bibnamefont {Kin-Lic~Chan}},\ }\href
  {https://pubs.acs.org/doi/10.1021/acs.chemrev.9b00829} {\bibfield  {journal}
  {\bibinfo  {journal} {Chem. Rev}\ }\textbf {\bibinfo {volume} {120}},\
  \bibinfo {pages} {12685} (\bibinfo {year} {2020})}\BibitemShut {NoStop}%
\bibitem [{\citenamefont {Motta}\ and\ \citenamefont
  {Rice}(2021)}]{motta2021emerging}%
  \BibitemOpen
  \bibfield  {author} {\bibinfo {author} {\bibfnamefont {M.}~\bibnamefont
  {Motta}}\ and\ \bibinfo {author} {\bibfnamefont {J.~E.}\ \bibnamefont
  {Rice}},\ }\href
  {https://wires.onlinelibrary.wiley.com/doi/abs/10.1002/wcms.1580} {\bibfield
  {journal} {\bibinfo  {journal} {WIREs Comput. Mol. Sci}\ ,\ \bibinfo {pages}
  {e1580}} (\bibinfo {year} {2021})}\BibitemShut {NoStop}%
\bibitem [{\citenamefont {Lloyd}(1996)}]{lloyd1996universal}%
  \BibitemOpen
  \bibfield  {author} {\bibinfo {author} {\bibfnamefont {S.}~\bibnamefont
  {Lloyd}},\ }\href {https://doi.org/10.1126/science.273.5278.1073} {\bibfield
  {journal} {\bibinfo  {journal} {Science}\ }\textbf {\bibinfo {volume}
  {273}},\ \bibinfo {pages} {1073} (\bibinfo {year} {1996})}\BibitemShut
  {NoStop}%
\bibitem [{\citenamefont {Martyn}\ \emph {et~al.}(2021)\citenamefont {Martyn},
  \citenamefont {Rossi}, \citenamefont {Tan},\ and\ \citenamefont
  {Chuang}}]{martyn2021grand}%
  \BibitemOpen
  \bibfield  {author} {\bibinfo {author} {\bibfnamefont {J.~M.}\ \bibnamefont
  {Martyn}}, \bibinfo {author} {\bibfnamefont {Z.~M.}\ \bibnamefont {Rossi}},
  \bibinfo {author} {\bibfnamefont {A.~K.}\ \bibnamefont {Tan}},\ and\ \bibinfo
  {author} {\bibfnamefont {I.~L.}\ \bibnamefont {Chuang}},\ }\href
  {https://journals.aps.org/prxquantum/abstract/10.1103/PRXQuantum.2.040203}
  {\bibfield  {journal} {\bibinfo  {journal} {PRX Quantum}\ }\textbf {\bibinfo
  {volume} {2}},\ \bibinfo {pages} {040203} (\bibinfo {year}
  {2021})}\BibitemShut {NoStop}%
\bibitem [{\citenamefont {McClean}\ \emph {et~al.}(2017)\citenamefont
  {McClean}, \citenamefont {Kimchi-Schwartz}, \citenamefont {Carter},\ and\
  \citenamefont {de~Jong}}]{mcclean2017subspace}%
  \BibitemOpen
  \bibfield  {author} {\bibinfo {author} {\bibfnamefont {J.~R.}\ \bibnamefont
  {McClean}}, \bibinfo {author} {\bibfnamefont {M.~E.}\ \bibnamefont
  {Kimchi-Schwartz}}, \bibinfo {author} {\bibfnamefont {J.}~\bibnamefont
  {Carter}},\ and\ \bibinfo {author} {\bibfnamefont {W.~A.}\ \bibnamefont
  {de~Jong}},\ }\href {https://doi.org/10.1103/PhysRevA.95.042308} {\bibfield
  {journal} {\bibinfo  {journal} {Phys. Rev. A}\ }\textbf {\bibinfo {volume}
  {95}},\ \bibinfo {pages} {042308} (\bibinfo {year} {2017})}\BibitemShut
  {NoStop}%
\bibitem [{\citenamefont {Takeshita}\ \emph
  {et~al.}(2020{\natexlab{a}})\citenamefont {Takeshita}, \citenamefont {Rubin},
  \citenamefont {Jiang}, \citenamefont {Lee}, \citenamefont {Babbush},\ and\
  \citenamefont {McClean}}]{takeshita2020subspace}%
  \BibitemOpen
  \bibfield  {author} {\bibinfo {author} {\bibfnamefont {T.}~\bibnamefont
  {Takeshita}}, \bibinfo {author} {\bibfnamefont {N.~C.}\ \bibnamefont
  {Rubin}}, \bibinfo {author} {\bibfnamefont {Z.}~\bibnamefont {Jiang}},
  \bibinfo {author} {\bibfnamefont {E.}~\bibnamefont {Lee}}, \bibinfo {author}
  {\bibfnamefont {R.}~\bibnamefont {Babbush}},\ and\ \bibinfo {author}
  {\bibfnamefont {J.~R.}\ \bibnamefont {McClean}},\ }\href
  {https://doi.org/10.1103/PhysRevX.10.011004} {\bibfield  {journal} {\bibinfo
  {journal} {Phys. Rev. X}\ }\textbf {\bibinfo {volume} {10}},\ \bibinfo
  {pages} {011004} (\bibinfo {year} {2020}{\natexlab{a}})}\BibitemShut
  {NoStop}%
\bibitem [{\citenamefont {Cohn}\ \emph {et~al.}(2021)\citenamefont {Cohn},
  \citenamefont {Motta},\ and\ \citenamefont
  {Parrish}}]{cohn2021filterdiagonalization}%
  \BibitemOpen
  \bibfield  {author} {\bibinfo {author} {\bibfnamefont {J.}~\bibnamefont
  {Cohn}}, \bibinfo {author} {\bibfnamefont {M.}~\bibnamefont {Motta}},\ and\
  \bibinfo {author} {\bibfnamefont {R.~M.}\ \bibnamefont {Parrish}},\ }\href
  {https://doi.org/10.1103/PRXQuantum.2.040352} {\bibfield  {journal} {\bibinfo
   {journal} {PRX Quantum}\ }\textbf {\bibinfo {volume} {2}},\ \bibinfo {pages}
  {040352} (\bibinfo {year} {2021})}\BibitemShut {NoStop}%
\bibitem [{\citenamefont {Yoshioka}\ \emph {et~al.}(2022)\citenamefont
  {Yoshioka}, \citenamefont {Hakoshima}, \citenamefont {Matsuzaki},
  \citenamefont {Tokunaga}, \citenamefont {Suzuki},\ and\ \citenamefont
  {Endo}}]{yoshioka2021virtualsubspace}%
  \BibitemOpen
  \bibfield  {author} {\bibinfo {author} {\bibfnamefont {N.}~\bibnamefont
  {Yoshioka}}, \bibinfo {author} {\bibfnamefont {H.}~\bibnamefont {Hakoshima}},
  \bibinfo {author} {\bibfnamefont {Y.}~\bibnamefont {Matsuzaki}}, \bibinfo
  {author} {\bibfnamefont {Y.}~\bibnamefont {Tokunaga}}, \bibinfo {author}
  {\bibfnamefont {Y.}~\bibnamefont {Suzuki}},\ and\ \bibinfo {author}
  {\bibfnamefont {S.}~\bibnamefont {Endo}},\ }\href
  {https://journals.aps.org/prl/abstract/10.1103/PhysRevLett.129.020502}
  {\bibfield  {journal} {\bibinfo  {journal} {Phys. Rev. Lett}\ }\textbf
  {\bibinfo {volume} {129}},\ \bibinfo {pages} {020502} (\bibinfo {year}
  {2022})}\BibitemShut {NoStop}%
\bibitem [{\citenamefont {Epperly}\ \emph {et~al.}(2021)\citenamefont
  {Epperly}, \citenamefont {Lin},\ and\ \citenamefont
  {Nakatsukasa}}]{epperly2021subspacediagonalization}%
  \BibitemOpen
  \bibfield  {author} {\bibinfo {author} {\bibfnamefont {E.~N.}\ \bibnamefont
  {Epperly}}, \bibinfo {author} {\bibfnamefont {L.}~\bibnamefont {Lin}},\ and\
  \bibinfo {author} {\bibfnamefont {Y.}~\bibnamefont {Nakatsukasa}},\ }\href
  {https://arxiv.org/abs/2110.07492} {\bibfield  {journal} {\bibinfo  {journal}
  {arXiv:2110.07492}\ } (\bibinfo {year} {2021})}\BibitemShut {NoStop}%
\bibitem [{\citenamefont {Baek}\ \emph {et~al.}(2022)\citenamefont {Baek},
  \citenamefont {Hait}, \citenamefont {Shee}, \citenamefont {Leimkuhler},
  \citenamefont {Huggins}, \citenamefont {Stetina}, \citenamefont
  {Head-Gordon},\ and\ \citenamefont {Whaley}}]{baek2022say}%
  \BibitemOpen
  \bibfield  {author} {\bibinfo {author} {\bibfnamefont {U.}~\bibnamefont
  {Baek}}, \bibinfo {author} {\bibfnamefont {D.}~\bibnamefont {Hait}}, \bibinfo
  {author} {\bibfnamefont {J.}~\bibnamefont {Shee}}, \bibinfo {author}
  {\bibfnamefont {O.}~\bibnamefont {Leimkuhler}}, \bibinfo {author}
  {\bibfnamefont {W.~J.}\ \bibnamefont {Huggins}}, \bibinfo {author}
  {\bibfnamefont {T.~F.}\ \bibnamefont {Stetina}}, \bibinfo {author}
  {\bibfnamefont {M.}~\bibnamefont {Head-Gordon}},\ and\ \bibinfo {author}
  {\bibfnamefont {K.~B.}\ \bibnamefont {Whaley}},\ }\href
  {https://arxiv.org/abs/2205.09039v1} {\bibfield  {journal} {\bibinfo
  {journal} {arXiv:2205.09039}\ } (\bibinfo {year} {2022})}\BibitemShut
  {NoStop}%
\bibitem [{\citenamefont {Colless}\ \emph {et~al.}(2018)\citenamefont
  {Colless}, \citenamefont {Ramasesh}, \citenamefont {Dahlen}, \citenamefont
  {Blok}, \citenamefont {Kimchi-Schwartz}, \citenamefont {McClean},
  \citenamefont {Carter}, \citenamefont {de~Jong},\ and\ \citenamefont
  {Siddiqi}}]{colless2018computation}%
  \BibitemOpen
  \bibfield  {author} {\bibinfo {author} {\bibfnamefont {J.~I.}\ \bibnamefont
  {Colless}}, \bibinfo {author} {\bibfnamefont {V.~V.}\ \bibnamefont
  {Ramasesh}}, \bibinfo {author} {\bibfnamefont {D.}~\bibnamefont {Dahlen}},
  \bibinfo {author} {\bibfnamefont {M.~S.}\ \bibnamefont {Blok}}, \bibinfo
  {author} {\bibfnamefont {M.~E.}\ \bibnamefont {Kimchi-Schwartz}}, \bibinfo
  {author} {\bibfnamefont {J.~R.}\ \bibnamefont {McClean}}, \bibinfo {author}
  {\bibfnamefont {J.}~\bibnamefont {Carter}}, \bibinfo {author} {\bibfnamefont
  {W.~A.}\ \bibnamefont {de~Jong}},\ and\ \bibinfo {author} {\bibfnamefont
  {I.}~\bibnamefont {Siddiqi}},\ }\href
  {https://journals.aps.org/prx/abstract/10.1103/PhysRevX.8.011021} {\bibfield
  {journal} {\bibinfo  {journal} {Phys. Rev. X}\ }\textbf {\bibinfo {volume}
  {8}},\ \bibinfo {pages} {011021} (\bibinfo {year} {2018})}\BibitemShut
  {NoStop}%
\bibitem [{\citenamefont {Eddins}\ \emph {et~al.}(2022)\citenamefont {Eddins},
  \citenamefont {Motta}, \citenamefont {Gujarati}, \citenamefont {Bravyi},
  \citenamefont {Mezzacapo}, \citenamefont {Hadfield},\ and\ \citenamefont
  {Sheldon}}]{eddins2021doubling}%
  \BibitemOpen
  \bibfield  {author} {\bibinfo {author} {\bibfnamefont {A.}~\bibnamefont
  {Eddins}}, \bibinfo {author} {\bibfnamefont {M.}~\bibnamefont {Motta}},
  \bibinfo {author} {\bibfnamefont {T.~P.}\ \bibnamefont {Gujarati}}, \bibinfo
  {author} {\bibfnamefont {S.}~\bibnamefont {Bravyi}}, \bibinfo {author}
  {\bibfnamefont {A.}~\bibnamefont {Mezzacapo}}, \bibinfo {author}
  {\bibfnamefont {C.}~\bibnamefont {Hadfield}},\ and\ \bibinfo {author}
  {\bibfnamefont {S.}~\bibnamefont {Sheldon}},\ }\href
  {https://doi.org/10.1103/PRXQuantum.3.010309} {\bibfield  {journal} {\bibinfo
   {journal} {PRX Quantum}\ }\textbf {\bibinfo {volume} {3}},\ \bibinfo {pages}
  {010309} (\bibinfo {year} {2022})}\BibitemShut {NoStop}%
\bibitem [{\citenamefont {Smart}\ and\ \citenamefont
  {Mazziotti}(2021)}]{smart2021quantum}%
  \BibitemOpen
  \bibfield  {author} {\bibinfo {author} {\bibfnamefont {S.~E.}\ \bibnamefont
  {Smart}}\ and\ \bibinfo {author} {\bibfnamefont {D.~A.}\ \bibnamefont
  {Mazziotti}},\ }\href
  {https://journals.aps.org/prl/abstract/10.1103/PhysRevLett.126.070504}
  {\bibfield  {journal} {\bibinfo  {journal} {Phys. Rev. Lett}\ }\textbf
  {\bibinfo {volume} {126}},\ \bibinfo {pages} {070504} (\bibinfo {year}
  {2021})}\BibitemShut {NoStop}%
\bibitem [{\citenamefont {Lax}(1952)}]{lax1952franck}%
  \BibitemOpen
  \bibfield  {author} {\bibinfo {author} {\bibfnamefont {M.}~\bibnamefont
  {Lax}},\ }\href {https://aip.scitation.org/doi/10.1063/1.1700283} {\bibfield
  {journal} {\bibinfo  {journal} {J. Chem. Phys}\ }\textbf {\bibinfo {volume}
  {20}},\ \bibinfo {pages} {1752} (\bibinfo {year} {1952})}\BibitemShut
  {NoStop}%
\bibitem [{\citenamefont {Heller}(1978)}]{heller1978photofragmentation}%
  \BibitemOpen
  \bibfield  {author} {\bibinfo {author} {\bibfnamefont {E.~J.}\ \bibnamefont
  {Heller}},\ }\href {https://aip.scitation.org/doi/10.1063/1.436197}
  {\bibfield  {journal} {\bibinfo  {journal} {J. Chem. Phys}\ }\textbf
  {\bibinfo {volume} {68}},\ \bibinfo {pages} {3891} (\bibinfo {year}
  {1978})}\BibitemShut {NoStop}%
\bibitem [{\citenamefont {Kulander}\ and\ \citenamefont
  {Heller}(1978)}]{kulander1978time}%
  \BibitemOpen
  \bibfield  {author} {\bibinfo {author} {\bibfnamefont {K.~C.}\ \bibnamefont
  {Kulander}}\ and\ \bibinfo {author} {\bibfnamefont {E.~J.}\ \bibnamefont
  {Heller}},\ }\href {https://aip.scitation.org/doi/10.1063/1.436930}
  {\bibfield  {journal} {\bibinfo  {journal} {J. Chem. Phys}\ }\textbf
  {\bibinfo {volume} {69}},\ \bibinfo {pages} {2439} (\bibinfo {year}
  {1978})}\BibitemShut {NoStop}%
\bibitem [{\citenamefont {Johnson}\ and\ \citenamefont
  {Kinsey}(1989)}]{johnson1989recurrences}%
  \BibitemOpen
  \bibfield  {author} {\bibinfo {author} {\bibfnamefont {B.~R.}\ \bibnamefont
  {Johnson}}\ and\ \bibinfo {author} {\bibfnamefont {J.~L.}\ \bibnamefont
  {Kinsey}},\ }\href {https://aip.scitation.org/doi/10.1063/1.457234}
  {\bibfield  {journal} {\bibinfo  {journal} {J. Chem. Phys}\ }\textbf
  {\bibinfo {volume} {91}},\ \bibinfo {pages} {7638} (\bibinfo {year}
  {1989})}\BibitemShut {NoStop}%
\bibitem [{\citenamefont {Gordon}(1965)}]{gordon1965molecular}%
  \BibitemOpen
  \bibfield  {author} {\bibinfo {author} {\bibfnamefont {R.}~\bibnamefont
  {Gordon}},\ }\href {https://aip.scitation.org/doi/10.1063/1.1696920}
  {\bibfield  {journal} {\bibinfo  {journal} {J. Chem. Phys}\ }\textbf
  {\bibinfo {volume} {43}},\ \bibinfo {pages} {1307} (\bibinfo {year}
  {1965})}\BibitemShut {NoStop}%
\bibitem [{\citenamefont {Boulet}\ and\ \citenamefont
  {Robert}(1982)}]{boulet1982short}%
  \BibitemOpen
  \bibfield  {author} {\bibinfo {author} {\bibfnamefont {C.}~\bibnamefont
  {Boulet}}\ and\ \bibinfo {author} {\bibfnamefont {D.}~\bibnamefont
  {Robert}},\ }\href {https://aip.scitation.org/doi/10.1063/1.444430}
  {\bibfield  {journal} {\bibinfo  {journal} {J. Chem. Phys}\ }\textbf
  {\bibinfo {volume} {77}},\ \bibinfo {pages} {4288} (\bibinfo {year}
  {1982})}\BibitemShut {NoStop}%
\bibitem [{\citenamefont {Clerk}\ \emph {et~al.}(2010)\citenamefont {Clerk},
  \citenamefont {Devoret}, \citenamefont {Girvin}, \citenamefont {Marquardt},\
  and\ \citenamefont {Schoelkopf}}]{clerk2010introduction}%
  \BibitemOpen
  \bibfield  {author} {\bibinfo {author} {\bibfnamefont {A.~A.}\ \bibnamefont
  {Clerk}}, \bibinfo {author} {\bibfnamefont {M.~H.}\ \bibnamefont {Devoret}},
  \bibinfo {author} {\bibfnamefont {S.~M.}\ \bibnamefont {Girvin}}, \bibinfo
  {author} {\bibfnamefont {F.}~\bibnamefont {Marquardt}},\ and\ \bibinfo
  {author} {\bibfnamefont {R.~J.}\ \bibnamefont {Schoelkopf}},\ }\href
  {https://journals.aps.org/rmp/pdf/10.1103/RevModPhys.82.1155} {\bibfield
  {journal} {\bibinfo  {journal} {Rev. Mod. Phys}\ }\textbf {\bibinfo {volume}
  {82}},\ \bibinfo {pages} {1155} (\bibinfo {year} {2010})}\BibitemShut
  {NoStop}%
\bibitem [{\citenamefont {Vitale}\ \emph {et~al.}(2015)\citenamefont {Vitale},
  \citenamefont {Dziedzic}, \citenamefont {Dubois}, \citenamefont {Fangohr},\
  and\ \citenamefont {Skylaris}}]{vitale2015anharmonic}%
  \BibitemOpen
  \bibfield  {author} {\bibinfo {author} {\bibfnamefont {V.}~\bibnamefont
  {Vitale}}, \bibinfo {author} {\bibfnamefont {J.}~\bibnamefont {Dziedzic}},
  \bibinfo {author} {\bibfnamefont {S.~M.-M.}\ \bibnamefont {Dubois}}, \bibinfo
  {author} {\bibfnamefont {H.}~\bibnamefont {Fangohr}},\ and\ \bibinfo {author}
  {\bibfnamefont {C.-K.}\ \bibnamefont {Skylaris}},\ }\href
  {https://pubs.acs.org/doi/10.1021/acs.jctc.5b00391} {\bibfield  {journal}
  {\bibinfo  {journal} {J. Chem. Theory Comput}\ }\textbf {\bibinfo {volume}
  {11}},\ \bibinfo {pages} {3321} (\bibinfo {year} {2015})}\BibitemShut
  {NoStop}%
\bibitem [{\citenamefont {Nascimento}\ and\ \citenamefont
  {DePrince~III}(2016)}]{nascimento2016linear}%
  \BibitemOpen
  \bibfield  {author} {\bibinfo {author} {\bibfnamefont {D.~R.}\ \bibnamefont
  {Nascimento}}\ and\ \bibinfo {author} {\bibfnamefont {A.~E.}\ \bibnamefont
  {DePrince~III}},\ }\href {https://pubs.acs.org/doi/10.1021/acs.jctc.6b00796}
  {\bibfield  {journal} {\bibinfo  {journal} {J. Chem. Theory Comput}\ }\textbf
  {\bibinfo {volume} {12}},\ \bibinfo {pages} {5834} (\bibinfo {year}
  {2016})}\BibitemShut {NoStop}%
\bibitem [{\citenamefont {Goings}\ \emph {et~al.}(2018)\citenamefont {Goings},
  \citenamefont {Lestrange},\ and\ \citenamefont {Li}}]{goings2018real}%
  \BibitemOpen
  \bibfield  {author} {\bibinfo {author} {\bibfnamefont {J.~J.}\ \bibnamefont
  {Goings}}, \bibinfo {author} {\bibfnamefont {P.~J.}\ \bibnamefont
  {Lestrange}},\ and\ \bibinfo {author} {\bibfnamefont {X.}~\bibnamefont
  {Li}},\ }\href {https://pubs.acs.org/doi/10.1021/acs.chemrev.0c00223}
  {\bibfield  {journal} {\bibinfo  {journal} {WIREs Comput. Mol. Sci}\ }\textbf
  {\bibinfo {volume} {8}},\ \bibinfo {pages} {e1341} (\bibinfo {year}
  {2018})}\BibitemShut {NoStop}%
\bibitem [{\citenamefont {Li}\ \emph {et~al.}(2020)\citenamefont {Li},
  \citenamefont {Govind}, \citenamefont {Isborn}, \citenamefont
  {DePrince~III},\ and\ \citenamefont {Lopata}}]{li2020real}%
  \BibitemOpen
  \bibfield  {author} {\bibinfo {author} {\bibfnamefont {X.}~\bibnamefont
  {Li}}, \bibinfo {author} {\bibfnamefont {N.}~\bibnamefont {Govind}}, \bibinfo
  {author} {\bibfnamefont {C.}~\bibnamefont {Isborn}}, \bibinfo {author}
  {\bibfnamefont {A.~E.}\ \bibnamefont {DePrince~III}},\ and\ \bibinfo {author}
  {\bibfnamefont {K.}~\bibnamefont {Lopata}},\ }\href
  {https://pubs.acs.org/doi/full/10.1021/acs.chemrev.0c00223} {\bibfield
  {journal} {\bibinfo  {journal} {Chem. Rev}\ }\textbf {\bibinfo {volume}
  {120}},\ \bibinfo {pages} {9951} (\bibinfo {year} {2020})}\BibitemShut
  {NoStop}%
\bibitem [{\citenamefont {Endo}\ \emph {et~al.}(2020)\citenamefont {Endo},
  \citenamefont {Kurata},\ and\ \citenamefont
  {Nakagawa}}]{endo2020calculation}%
  \BibitemOpen
  \bibfield  {author} {\bibinfo {author} {\bibfnamefont {S.}~\bibnamefont
  {Endo}}, \bibinfo {author} {\bibfnamefont {I.}~\bibnamefont {Kurata}},\ and\
  \bibinfo {author} {\bibfnamefont {Y.~O.}\ \bibnamefont {Nakagawa}},\ }\href
  {https://journals.aps.org/prresearch/abstract/10.1103/PhysRevResearch.2.033281}
  {\bibfield  {journal} {\bibinfo  {journal} {Phys. Rev. Research}\ }\textbf
  {\bibinfo {volume} {2}},\ \bibinfo {pages} {033281} (\bibinfo {year}
  {2020})}\BibitemShut {NoStop}%
\bibitem [{\citenamefont {Bravyi}\ \emph {et~al.}(2016)\citenamefont {Bravyi},
  \citenamefont {Smith},\ and\ \citenamefont {Smolin}}]{bravyi2016trading}%
  \BibitemOpen
  \bibfield  {author} {\bibinfo {author} {\bibfnamefont {S.}~\bibnamefont
  {Bravyi}}, \bibinfo {author} {\bibfnamefont {G.}~\bibnamefont {Smith}},\ and\
  \bibinfo {author} {\bibfnamefont {J.~A.}\ \bibnamefont {Smolin}},\ }\href
  {https://journals.aps.org/prx/abstract/10.1103/PhysRevX.6.021043} {\bibfield
  {journal} {\bibinfo  {journal} {Phys. Rev. X}\ }\textbf {\bibinfo {volume}
  {6}},\ \bibinfo {pages} {021043} (\bibinfo {year} {2016})}\BibitemShut
  {NoStop}%
\bibitem [{\citenamefont {Huembeli}\ \emph {et~al.}(2022)\citenamefont
  {Huembeli}, \citenamefont {Carleo},\ and\ \citenamefont
  {Mezzacapo}}]{huembeli2022entanglement}%
  \BibitemOpen
  \bibfield  {author} {\bibinfo {author} {\bibfnamefont {P.}~\bibnamefont
  {Huembeli}}, \bibinfo {author} {\bibfnamefont {G.}~\bibnamefont {Carleo}},\
  and\ \bibinfo {author} {\bibfnamefont {A.}~\bibnamefont {Mezzacapo}},\ }\href
  {https://arxiv.org/abs/2205.00933v1} {\bibfield  {journal} {\bibinfo
  {journal} {arXiv:2205.00933}\ } (\bibinfo {year} {2022})}\BibitemShut
  {NoStop}%
\bibitem [{\citenamefont {Aleksandrowicz}\ \emph {et~al.}(2019)\citenamefont
  {Aleksandrowicz}, \citenamefont {Alexander}, \citenamefont {Barkoutsos},
  \citenamefont {Bello}, \citenamefont {Ben-Haim}, \citenamefont {Bucher},
  \citenamefont {Cabrera-Hern{\'a}ndez}, \citenamefont {Carballo-Franquis},
  \citenamefont {Chen}, \citenamefont {Chen} \emph
  {et~al.}}]{aleksandrowicz2019qiskit}%
  \BibitemOpen
  \bibfield  {author} {\bibinfo {author} {\bibfnamefont {G.}~\bibnamefont
  {Aleksandrowicz}}, \bibinfo {author} {\bibfnamefont {T.}~\bibnamefont
  {Alexander}}, \bibinfo {author} {\bibfnamefont {P.}~\bibnamefont
  {Barkoutsos}}, \bibinfo {author} {\bibfnamefont {L.}~\bibnamefont {Bello}},
  \bibinfo {author} {\bibfnamefont {Y.}~\bibnamefont {Ben-Haim}}, \bibinfo
  {author} {\bibfnamefont {D.}~\bibnamefont {Bucher}}, \bibinfo {author}
  {\bibfnamefont {F.}~\bibnamefont {Cabrera-Hern{\'a}ndez}}, \bibinfo {author}
  {\bibfnamefont {J.}~\bibnamefont {Carballo-Franquis}}, \bibinfo {author}
  {\bibfnamefont {A.}~\bibnamefont {Chen}}, \bibinfo {author} {\bibfnamefont
  {C.}~\bibnamefont {Chen}}, \emph {et~al.},\ }\href
  {https://zenodo.org/record/2562111#.XhA8qi2ZPyI} {\bibinfo {title} {Qiskit:
  An open-source framework for quantum computing}} (\bibinfo {year}
  {2019})\BibitemShut {NoStop}%
\bibitem [{\citenamefont {Sun}\ \emph {et~al.}(2018)\citenamefont {Sun},
  \citenamefont {Berkelbach}, \citenamefont {Blunt}, \citenamefont {Booth},
  \citenamefont {Guo}, \citenamefont {Li}, \citenamefont {Liu}, \citenamefont
  {McClain}, \citenamefont {Sayfutyarova}, \citenamefont {Sharma} \emph
  {et~al.}}]{sun2018pyscf}%
  \BibitemOpen
  \bibfield  {author} {\bibinfo {author} {\bibfnamefont {Q.}~\bibnamefont
  {Sun}}, \bibinfo {author} {\bibfnamefont {T.~C.}\ \bibnamefont {Berkelbach}},
  \bibinfo {author} {\bibfnamefont {N.~S.}\ \bibnamefont {Blunt}}, \bibinfo
  {author} {\bibfnamefont {G.~H.}\ \bibnamefont {Booth}}, \bibinfo {author}
  {\bibfnamefont {S.}~\bibnamefont {Guo}}, \bibinfo {author} {\bibfnamefont
  {Z.}~\bibnamefont {Li}}, \bibinfo {author} {\bibfnamefont {J.}~\bibnamefont
  {Liu}}, \bibinfo {author} {\bibfnamefont {J.~D.}\ \bibnamefont {McClain}},
  \bibinfo {author} {\bibfnamefont {E.~R.}\ \bibnamefont {Sayfutyarova}},
  \bibinfo {author} {\bibfnamefont {S.}~\bibnamefont {Sharma}}, \emph
  {et~al.},\ }\href {https://onlinelibrary.wiley.com/doi/abs/10.1002/wcms.1340}
  {\bibfield  {journal} {\bibinfo  {journal} {WIREs Comput. Mol. Sci}\ }\textbf
  {\bibinfo {volume} {8}},\ \bibinfo {pages} {e1340} (\bibinfo {year}
  {2018})}\BibitemShut {NoStop}%
\bibitem [{\citenamefont {Sun}\ \emph {et~al.}(2020)\citenamefont {Sun} \emph
  {et~al.}}]{sun2020recent}%
  \BibitemOpen
  \bibfield  {author} {\bibinfo {author} {\bibfnamefont {Q.}~\bibnamefont
  {Sun}} \emph {et~al.},\ }\href {https://doi.org/10.1063/5.0006074} {\bibfield
   {journal} {\bibinfo  {journal} {J. Chem. Phys}\ }\textbf {\bibinfo {volume}
  {153}},\ \bibinfo {pages} {024109} (\bibinfo {year} {2020})}\BibitemShut
  {NoStop}%
\bibitem [{\citenamefont {Maciejewski}\ \emph {et~al.}(2020)\citenamefont
  {Maciejewski}, \citenamefont {Zimbor{\'a}s},\ and\ \citenamefont
  {Oszmaniec}}]{maciejewski2020mitigation}%
  \BibitemOpen
  \bibfield  {author} {\bibinfo {author} {\bibfnamefont {F.~B.}\ \bibnamefont
  {Maciejewski}}, \bibinfo {author} {\bibfnamefont {Z.}~\bibnamefont
  {Zimbor{\'a}s}},\ and\ \bibinfo {author} {\bibfnamefont {M.}~\bibnamefont
  {Oszmaniec}},\ }\href {https://quantum-journal.org/papers/q-2020-04-24-257/}
  {\bibfield  {journal} {\bibinfo  {journal} {Quantum}\ }\textbf {\bibinfo
  {volume} {4}},\ \bibinfo {pages} {257} (\bibinfo {year} {2020})}\BibitemShut
  {NoStop}%
\bibitem [{\citenamefont {Nation}\ \emph {et~al.}(2021)\citenamefont {Nation},
  \citenamefont {Kang}, \citenamefont {Sundaresan},\ and\ \citenamefont
  {Gambetta}}]{nation2021scalable}%
  \BibitemOpen
  \bibfield  {author} {\bibinfo {author} {\bibfnamefont {P.~D.}\ \bibnamefont
  {Nation}}, \bibinfo {author} {\bibfnamefont {H.}~\bibnamefont {Kang}},
  \bibinfo {author} {\bibfnamefont {N.}~\bibnamefont {Sundaresan}},\ and\
  \bibinfo {author} {\bibfnamefont {J.~M.}\ \bibnamefont {Gambetta}},\ }\href
  {https://journals.aps.org/prxquantum/abstract/10.1103/PRXQuantum.2.040326}
  {\bibfield  {journal} {\bibinfo  {journal} {PRX Quantum}\ }\textbf {\bibinfo
  {volume} {2}},\ \bibinfo {pages} {040326} (\bibinfo {year}
  {2021})}\BibitemShut {NoStop}%
\bibitem [{\citenamefont {Huggins}\ \emph {et~al.}(2021)\citenamefont
  {Huggins}, \citenamefont {McClean}, \citenamefont {Rubin}, \citenamefont
  {Jiang}, \citenamefont {Wiebe}, \citenamefont {Whaley},\ and\ \citenamefont
  {Babbush}}]{huggins2021efficient}%
  \BibitemOpen
  \bibfield  {author} {\bibinfo {author} {\bibfnamefont {W.~J.}\ \bibnamefont
  {Huggins}}, \bibinfo {author} {\bibfnamefont {J.~R.}\ \bibnamefont
  {McClean}}, \bibinfo {author} {\bibfnamefont {N.~C.}\ \bibnamefont {Rubin}},
  \bibinfo {author} {\bibfnamefont {Z.}~\bibnamefont {Jiang}}, \bibinfo
  {author} {\bibfnamefont {N.}~\bibnamefont {Wiebe}}, \bibinfo {author}
  {\bibfnamefont {K.~B.}\ \bibnamefont {Whaley}},\ and\ \bibinfo {author}
  {\bibfnamefont {R.}~\bibnamefont {Babbush}},\ }\href
  {https://www.nature.com/articles/s41534-020-00341-7} {\bibfield  {journal}
  {\bibinfo  {journal} {npj Quantum Inf}\ }\textbf {\bibinfo {volume} {7}},\
  \bibinfo {pages} {1} (\bibinfo {year} {2021})}\BibitemShut {NoStop}%
\bibitem [{\citenamefont {Gottesman}(1998)}]{gottesman1998heisenberg}%
  \BibitemOpen
  \bibfield  {author} {\bibinfo {author} {\bibfnamefont {D.}~\bibnamefont
  {Gottesman}},\ }\href {https://arxiv.org/abs/quant-ph/9807006} {\bibfield
  {journal} {\bibinfo  {journal} {quant-ph/9807006}\ } (\bibinfo {year}
  {1998})}\BibitemShut {NoStop}%
\bibitem [{\citenamefont {Czarnik}\ \emph {et~al.}(2021)\citenamefont
  {Czarnik}, \citenamefont {Arrasmith}, \citenamefont {Coles},\ and\
  \citenamefont {Cincio}}]{czarnik2021error}%
  \BibitemOpen
  \bibfield  {author} {\bibinfo {author} {\bibfnamefont {P.}~\bibnamefont
  {Czarnik}}, \bibinfo {author} {\bibfnamefont {A.}~\bibnamefont {Arrasmith}},
  \bibinfo {author} {\bibfnamefont {P.~J.}\ \bibnamefont {Coles}},\ and\
  \bibinfo {author} {\bibfnamefont {L.}~\bibnamefont {Cincio}},\ }\href
  {https://quantum-journal.org/papers/q-2021-11-26-592/} {\bibfield  {journal}
  {\bibinfo  {journal} {Quantum}\ }\textbf {\bibinfo {volume} {5}},\ \bibinfo
  {pages} {592} (\bibinfo {year} {2021})}\BibitemShut {NoStop}%
\bibitem [{\citenamefont {Nielsen}\ and\ \citenamefont
  {Chuang}(2010)}]{nielsen2002quantum}%
  \BibitemOpen
  \bibfield  {author} {\bibinfo {author} {\bibfnamefont {M.~A.}\ \bibnamefont
  {Nielsen}}\ and\ \bibinfo {author} {\bibfnamefont {I.~L.}\ \bibnamefont
  {Chuang}},\ }\href@noop {} {\emph {\bibinfo {title} {Quantum computation and
  quantum information}}}\ (\bibinfo  {publisher} {Cambridge University Press},\
  \bibinfo {year} {2010})\BibitemShut {NoStop}%
\bibitem [{\citenamefont {Johnson~III}\ \emph {et~al.}(1999)\citenamefont
  {Johnson~III} \emph {et~al.}}]{johnson1999nist}%
  \BibitemOpen
  \bibfield  {author} {\bibinfo {author} {\bibfnamefont {R.~D.}\ \bibnamefont
  {Johnson~III}} \emph {et~al.},\ }\href {https://cccbdb.nist.gov} {\bibinfo
  {title} {Nist 101. {C}omputational chemistry comparison and benchmark
  database}} (\bibinfo {year} {1999})\BibitemShut {NoStop}%
\bibitem [{Note1()}]{Note1}%
  \BibitemOpen
  \bibinfo {note} {Since $\protect \mathrm {E}(\ce {H2S+ + H})$ - $\protect
  \mathrm {E}(\ce {H2S + H+})$ = $\protect \mathrm {[ E(H_2S^+)}$ - $\protect
  \mathrm {E(H_2S) ]}$ + $\protect \mathrm {[ E(H)}$ - $\protect \mathrm
  {E(H^+) ]}$ = $\protect \mathrm {IP(H_2S)}$ - $\protect \mathrm {IP(H)}$, the
  inequality $\protect \mathrm {IP(H_2S)}$ - $\protect \mathrm {IP(H)<0}$
  implies $\protect \mathrm {E(H_2S^+ + H)}$ $<$ $\protect \mathrm {E(H_2S +
  H^+)}$.}\BibitemShut {Stop}%
\bibitem [{\citenamefont {Gao}\ \emph {et~al.}(2021)\citenamefont {Gao},
  \citenamefont {Jones}, \citenamefont {Motta}, \citenamefont {Sugawara},
  \citenamefont {Watanabe}, \citenamefont {Kobayashi}, \citenamefont
  {Watanabe}, \citenamefont {Ohnishi}, \citenamefont {Nakamura},\ and\
  \citenamefont {Yamamoto}}]{gao2021applications}%
  \BibitemOpen
  \bibfield  {author} {\bibinfo {author} {\bibfnamefont {Q.}~\bibnamefont
  {Gao}}, \bibinfo {author} {\bibfnamefont {G.~O.}\ \bibnamefont {Jones}},
  \bibinfo {author} {\bibfnamefont {M.}~\bibnamefont {Motta}}, \bibinfo
  {author} {\bibfnamefont {M.}~\bibnamefont {Sugawara}}, \bibinfo {author}
  {\bibfnamefont {H.~C.}\ \bibnamefont {Watanabe}}, \bibinfo {author}
  {\bibfnamefont {T.}~\bibnamefont {Kobayashi}}, \bibinfo {author}
  {\bibfnamefont {E.}~\bibnamefont {Watanabe}}, \bibinfo {author}
  {\bibfnamefont {Y.-y.}\ \bibnamefont {Ohnishi}}, \bibinfo {author}
  {\bibfnamefont {H.}~\bibnamefont {Nakamura}},\ and\ \bibinfo {author}
  {\bibfnamefont {N.}~\bibnamefont {Yamamoto}},\ }\href
  {https://www.nature.com/articles/s41524-021-00540-6} {\bibfield  {journal}
  {\bibinfo  {journal} {npj Comput. Mater}\ }\textbf {\bibinfo {volume} {7}},\
  \bibinfo {pages} {1} (\bibinfo {year} {2021})}\BibitemShut {NoStop}%
\bibitem [{\citenamefont {Huang}\ \emph {et~al.}(2022)\citenamefont {Huang},
  \citenamefont {Cai}, \citenamefont {Li}, \citenamefont {Ge}, \citenamefont
  {Hou}, \citenamefont {Li}, \citenamefont {Liu}, \citenamefont {Shi},
  \citenamefont {Chen}, \citenamefont {Zheng} \emph
  {et~al.}}]{huang2022variational}%
  \BibitemOpen
  \bibfield  {author} {\bibinfo {author} {\bibfnamefont {K.}~\bibnamefont
  {Huang}}, \bibinfo {author} {\bibfnamefont {X.}~\bibnamefont {Cai}}, \bibinfo
  {author} {\bibfnamefont {H.}~\bibnamefont {Li}}, \bibinfo {author}
  {\bibfnamefont {Z.-Y.}\ \bibnamefont {Ge}}, \bibinfo {author} {\bibfnamefont
  {R.}~\bibnamefont {Hou}}, \bibinfo {author} {\bibfnamefont {H.}~\bibnamefont
  {Li}}, \bibinfo {author} {\bibfnamefont {T.}~\bibnamefont {Liu}}, \bibinfo
  {author} {\bibfnamefont {Y.}~\bibnamefont {Shi}}, \bibinfo {author}
  {\bibfnamefont {C.}~\bibnamefont {Chen}}, \bibinfo {author} {\bibfnamefont
  {D.}~\bibnamefont {Zheng}}, \emph {et~al.},\ }\href@noop {} {\bibfield
  {journal} {\bibinfo  {journal} {The Journal of Physical Chemistry Letters}\
  }\textbf {\bibinfo {volume} {13}},\ \bibinfo {pages} {9114} (\bibinfo {year}
  {2022})}\BibitemShut {NoStop}%
\bibitem [{\citenamefont {Khan}\ \emph {et~al.}(2022)\citenamefont {Khan},
  \citenamefont {Tudorovskaya}, \citenamefont {Kirsopp}, \citenamefont {Ramo},
  \citenamefont {Warrier}, \citenamefont {Papanastasiou},\ and\ \citenamefont
  {Singh}}]{khan2022chemically}%
  \BibitemOpen
  \bibfield  {author} {\bibinfo {author} {\bibfnamefont {I.}~\bibnamefont
  {Khan}}, \bibinfo {author} {\bibfnamefont {M.}~\bibnamefont {Tudorovskaya}},
  \bibinfo {author} {\bibfnamefont {J.}~\bibnamefont {Kirsopp}}, \bibinfo
  {author} {\bibfnamefont {D.~M.}\ \bibnamefont {Ramo}}, \bibinfo {author}
  {\bibfnamefont {P.}~\bibnamefont {Warrier}}, \bibinfo {author} {\bibfnamefont
  {D.}~\bibnamefont {Papanastasiou}},\ and\ \bibinfo {author} {\bibfnamefont
  {R.}~\bibnamefont {Singh}},\ }\href@noop {} {\bibfield  {journal} {\bibinfo
  {journal} {arXiv:2210.14834}\ } (\bibinfo {year} {2022})}\BibitemShut
  {NoStop}%
\bibitem [{\citenamefont {Tammaro}\ \emph {et~al.}(2023)\citenamefont
  {Tammaro}, \citenamefont {Galli}, \citenamefont {Rice},\ and\ \citenamefont
  {Motta}}]{tammaro2022n}%
  \BibitemOpen
  \bibfield  {author} {\bibinfo {author} {\bibfnamefont {A.}~\bibnamefont
  {Tammaro}}, \bibinfo {author} {\bibfnamefont {D.~E.}\ \bibnamefont {Galli}},
  \bibinfo {author} {\bibfnamefont {J.~E.}\ \bibnamefont {Rice}},\ and\
  \bibinfo {author} {\bibfnamefont {M.}~\bibnamefont {Motta}},\ }\href
  {https://pubs.acs.org/doi/10.1021/acs.jpca.2c07653} {\bibfield  {journal}
  {\bibinfo  {journal} {J. Phys. Chem A}\ }\textbf {\bibinfo {volume} {127}},\
  \bibinfo {pages} {817} (\bibinfo {year} {2023})}\BibitemShut {NoStop}%
\bibitem [{\citenamefont {Takeshita}\ \emph
  {et~al.}(2020{\natexlab{b}})\citenamefont {Takeshita}, \citenamefont {Rubin},
  \citenamefont {Jiang}, \citenamefont {Lee}, \citenamefont {Babbush},\ and\
  \citenamefont {McClean}}]{takeshita2020increasing}%
  \BibitemOpen
  \bibfield  {author} {\bibinfo {author} {\bibfnamefont {T.}~\bibnamefont
  {Takeshita}}, \bibinfo {author} {\bibfnamefont {N.~C.}\ \bibnamefont
  {Rubin}}, \bibinfo {author} {\bibfnamefont {Z.}~\bibnamefont {Jiang}},
  \bibinfo {author} {\bibfnamefont {E.}~\bibnamefont {Lee}}, \bibinfo {author}
  {\bibfnamefont {R.}~\bibnamefont {Babbush}},\ and\ \bibinfo {author}
  {\bibfnamefont {J.~R.}\ \bibnamefont {McClean}},\ }\href
  {https://journals.aps.org/prx/abstract/10.1103/PhysRevX.10.011004} {\bibfield
   {journal} {\bibinfo  {journal} {Phys. Rev. X}\ }\textbf {\bibinfo {volume}
  {10}},\ \bibinfo {pages} {011004} (\bibinfo {year}
  {2020}{\natexlab{b}})}\BibitemShut {NoStop}%
\bibitem [{\citenamefont {Boyn}\ \emph {et~al.}(2021)\citenamefont {Boyn},
  \citenamefont {Lykhin}, \citenamefont {Smart}, \citenamefont {Gagliardi},\
  and\ \citenamefont {Mazziotti}}]{boyn2021quantum}%
  \BibitemOpen
  \bibfield  {author} {\bibinfo {author} {\bibfnamefont {J.-N.}\ \bibnamefont
  {Boyn}}, \bibinfo {author} {\bibfnamefont {A.~O.}\ \bibnamefont {Lykhin}},
  \bibinfo {author} {\bibfnamefont {S.~E.}\ \bibnamefont {Smart}}, \bibinfo
  {author} {\bibfnamefont {L.}~\bibnamefont {Gagliardi}},\ and\ \bibinfo
  {author} {\bibfnamefont {D.~A.}\ \bibnamefont {Mazziotti}},\ }\href
  {https://aip.scitation.org/doi/10.1063/5.0074842} {\bibfield  {journal}
  {\bibinfo  {journal} {J. Chem. Phys}\ }\textbf {\bibinfo {volume} {155}},\
  \bibinfo {pages} {244106} (\bibinfo {year} {2021})}\BibitemShut {NoStop}%
\bibitem [{\citenamefont {Nakano}\ and\ \citenamefont
  {Hirao}(2000)}]{nakano2000quasi}%
  \BibitemOpen
  \bibfield  {author} {\bibinfo {author} {\bibfnamefont {H.}~\bibnamefont
  {Nakano}}\ and\ \bibinfo {author} {\bibfnamefont {K.}~\bibnamefont {Hirao}},\
  }\href
  {https://www.sciencedirect.com/science/article/abs/pii/S0009261499013640}
  {\bibfield  {journal} {\bibinfo  {journal} {Chem. Phys. Lett}\ }\textbf
  {\bibinfo {volume} {317}},\ \bibinfo {pages} {90} (\bibinfo {year}
  {2000})}\BibitemShut {NoStop}%
\bibitem [{\citenamefont {Nakano}\ \emph {et~al.}(2001)\citenamefont {Nakano},
  \citenamefont {Nakatani},\ and\ \citenamefont {Hirao}}]{nakano2001second}%
  \BibitemOpen
  \bibfield  {author} {\bibinfo {author} {\bibfnamefont {H.}~\bibnamefont
  {Nakano}}, \bibinfo {author} {\bibfnamefont {J.}~\bibnamefont {Nakatani}},\
  and\ \bibinfo {author} {\bibfnamefont {K.}~\bibnamefont {Hirao}},\ }\href
  {https://aip.scitation.org/doi/abs/10.1063/1.1332992?cookieSet=1} {\bibfield
  {journal} {\bibinfo  {journal} {J. Chem. Phys}\ }\textbf {\bibinfo {volume}
  {114}},\ \bibinfo {pages} {1133} (\bibinfo {year} {2001})}\BibitemShut
  {NoStop}%
\bibitem [{\citenamefont {Kawashima}\ \emph {et~al.}(2021)\citenamefont
  {Kawashima}, \citenamefont {Lloyd}, \citenamefont {Coons}, \citenamefont
  {Nam}, \citenamefont {Matsuura}, \citenamefont {Garza}, \citenamefont
  {Johri}, \citenamefont {Huntington}, \citenamefont {Senicourt}, \citenamefont
  {Maksymov} \emph {et~al.}}]{kawashima2021optimizing}%
  \BibitemOpen
  \bibfield  {author} {\bibinfo {author} {\bibfnamefont {Y.}~\bibnamefont
  {Kawashima}}, \bibinfo {author} {\bibfnamefont {E.}~\bibnamefont {Lloyd}},
  \bibinfo {author} {\bibfnamefont {M.~P.}\ \bibnamefont {Coons}}, \bibinfo
  {author} {\bibfnamefont {Y.}~\bibnamefont {Nam}}, \bibinfo {author}
  {\bibfnamefont {S.}~\bibnamefont {Matsuura}}, \bibinfo {author}
  {\bibfnamefont {A.~J.}\ \bibnamefont {Garza}}, \bibinfo {author}
  {\bibfnamefont {S.}~\bibnamefont {Johri}}, \bibinfo {author} {\bibfnamefont
  {L.}~\bibnamefont {Huntington}}, \bibinfo {author} {\bibfnamefont
  {V.}~\bibnamefont {Senicourt}}, \bibinfo {author} {\bibfnamefont {A.~O.}\
  \bibnamefont {Maksymov}}, \emph {et~al.},\ }\href
  {https://www.nature.com/articles/s42005-021-00751-9} {\bibfield  {journal}
  {\bibinfo  {journal} {Comm. Phys}\ }\textbf {\bibinfo {volume} {4}},\
  \bibinfo {pages} {1} (\bibinfo {year} {2021})}\BibitemShut {NoStop}%
\bibitem [{\citenamefont {Ma}\ \emph {et~al.}(2020)\citenamefont {Ma},
  \citenamefont {Govoni},\ and\ \citenamefont {Galli}}]{ma2020quantum}%
  \BibitemOpen
  \bibfield  {author} {\bibinfo {author} {\bibfnamefont {H.}~\bibnamefont
  {Ma}}, \bibinfo {author} {\bibfnamefont {M.}~\bibnamefont {Govoni}},\ and\
  \bibinfo {author} {\bibfnamefont {G.}~\bibnamefont {Galli}},\ }\href
  {https://www.nature.com/articles/s41524-020-00353-z} {\bibfield  {journal}
  {\bibinfo  {journal} {npj Comput. Mater}\ }\textbf {\bibinfo {volume} {6}},\
  \bibinfo {pages} {1} (\bibinfo {year} {2020})}\BibitemShut {NoStop}%
\bibitem [{\citenamefont {Cheng}\ \emph {et~al.}(2020)\citenamefont {Cheng},
  \citenamefont {Deumens}, \citenamefont {Freericks}, \citenamefont {Li},\ and\
  \citenamefont {Sanders}}]{cheng2020application}%
  \BibitemOpen
  \bibfield  {author} {\bibinfo {author} {\bibfnamefont {H.-P.}\ \bibnamefont
  {Cheng}}, \bibinfo {author} {\bibfnamefont {E.}~\bibnamefont {Deumens}},
  \bibinfo {author} {\bibfnamefont {J.~K.}\ \bibnamefont {Freericks}}, \bibinfo
  {author} {\bibfnamefont {C.}~\bibnamefont {Li}},\ and\ \bibinfo {author}
  {\bibfnamefont {B.~A.}\ \bibnamefont {Sanders}},\ }\href
  {https://www.frontiersin.org/articles/10.3389/fchem.2020.587143/full}
  {\bibfield  {journal} {\bibinfo  {journal} {Front. Chem}\ }\textbf {\bibinfo
  {volume} {8}},\ \bibinfo {pages} {587143} (\bibinfo {year}
  {2020})}\BibitemShut {NoStop}%
\bibitem [{\citenamefont {Castaldo}\ \emph {et~al.}(2021)\citenamefont
  {Castaldo}, \citenamefont {Jahangiri}, \citenamefont {Delgado},\ and\
  \citenamefont {Corni}}]{castaldo2021quantum}%
  \BibitemOpen
  \bibfield  {author} {\bibinfo {author} {\bibfnamefont {D.}~\bibnamefont
  {Castaldo}}, \bibinfo {author} {\bibfnamefont {S.}~\bibnamefont {Jahangiri}},
  \bibinfo {author} {\bibfnamefont {A.}~\bibnamefont {Delgado}},\ and\ \bibinfo
  {author} {\bibfnamefont {S.}~\bibnamefont {Corni}},\ }\href
  {https://arxiv.org/abs/2111.13458} {\bibfield  {journal} {\bibinfo  {journal}
  {arXiv:2111.13458}\ } (\bibinfo {year} {2021})}\BibitemShut {NoStop}%
\bibitem [{\citenamefont {Beebe}\ and\ \citenamefont
  {Linderberg}(1977)}]{Beebe:1977:683}%
  \BibitemOpen
  \bibfield  {author} {\bibinfo {author} {\bibfnamefont {N.~H.~F.}\
  \bibnamefont {Beebe}}\ and\ \bibinfo {author} {\bibfnamefont
  {J.}~\bibnamefont {Linderberg}},\ }\href {https://doi.org/Two-electron
  Integrals in Molecular Calculations} {\bibfield  {journal} {\bibinfo
  {journal} {Int. J. Quantum Chem}\ }\textbf {\bibinfo {volume} {12}},\
  \bibinfo {pages} {683} (\bibinfo {year} {1977})}\BibitemShut {NoStop}%
\bibitem [{\citenamefont {Aquilante}\ \emph {et~al.}(2007)\citenamefont
  {Aquilante}, \citenamefont {Pedersen},\ and\ \citenamefont
  {Lindh}}]{Aquilante:2007:194106}%
  \BibitemOpen
  \bibfield  {author} {\bibinfo {author} {\bibfnamefont {F.}~\bibnamefont
  {Aquilante}}, \bibinfo {author} {\bibfnamefont {T.~B.}\ \bibnamefont
  {Pedersen}},\ and\ \bibinfo {author} {\bibfnamefont {R.}~\bibnamefont
  {Lindh}},\ }\href {https://doi.org/10.1063/1.2736701} {\bibfield  {journal}
  {\bibinfo  {journal} {J. Chem. Phys}\ }\textbf {\bibinfo {volume} {126}},\
  \bibinfo {pages} {194106} (\bibinfo {year} {2007})}\BibitemShut {NoStop}%
\bibitem [{\citenamefont {Motta}\ and\ \citenamefont
  {Zhang}(2018)}]{motta2018ab}%
  \BibitemOpen
  \bibfield  {author} {\bibinfo {author} {\bibfnamefont {M.}~\bibnamefont
  {Motta}}\ and\ \bibinfo {author} {\bibfnamefont {S.}~\bibnamefont {Zhang}},\
  }\href {https://onlinelibrary.wiley.com/doi/abs/10.1002/wcms.1364} {\bibfield
   {journal} {\bibinfo  {journal} {WIREs Comput. Mol. Sci}\ }\textbf {\bibinfo
  {volume} {8}},\ \bibinfo {pages} {e1364} (\bibinfo {year}
  {2018})}\BibitemShut {NoStop}%
\bibitem [{\citenamefont {Motta}\ \emph {et~al.}(2019)\citenamefont {Motta},
  \citenamefont {Shee}, \citenamefont {Zhang},\ and\ \citenamefont
  {Chan}}]{motta2019efficient}%
  \BibitemOpen
  \bibfield  {author} {\bibinfo {author} {\bibfnamefont {M.}~\bibnamefont
  {Motta}}, \bibinfo {author} {\bibfnamefont {J.}~\bibnamefont {Shee}},
  \bibinfo {author} {\bibfnamefont {S.}~\bibnamefont {Zhang}},\ and\ \bibinfo
  {author} {\bibfnamefont {G.~K.-L.}\ \bibnamefont {Chan}},\ }\href
  {https://pubs.acs.org/doi/abs/10.1021/acs.jctc.8b00996} {\bibfield  {journal}
  {\bibinfo  {journal} {J. Chem. Theory Comput}\ }\textbf {\bibinfo {volume}
  {15}},\ \bibinfo {pages} {3510} (\bibinfo {year} {2019})}\BibitemShut
  {NoStop}%
\bibitem [{\citenamefont {Bello}\ \emph {et~al.}(2021)\citenamefont {Bello},
  \citenamefont {Bra\'{n}czyk}, \citenamefont {Bravyi}, \citenamefont {Eddins},
  \citenamefont {Gacon}, \citenamefont {Gujarati}, \citenamefont {Hamamura},
  \citenamefont {Imamichi}, \citenamefont {Johnson}, \citenamefont
  {Liepuoniute}, \citenamefont {Motta}, \citenamefont {Rossmannek},
  \citenamefont {Scholten}, \citenamefont {Sitdikov},\ and\ \citenamefont
  {Woerner}}]{entanglement-forging}%
  \BibitemOpen
  \bibfield  {author} {\bibinfo {author} {\bibfnamefont {L.}~\bibnamefont
  {Bello}}, \bibinfo {author} {\bibfnamefont {A.~M.}\ \bibnamefont
  {Bra\'{n}czyk}}, \bibinfo {author} {\bibfnamefont {S.}~\bibnamefont
  {Bravyi}}, \bibinfo {author} {\bibfnamefont {A.}~\bibnamefont {Eddins}},
  \bibinfo {author} {\bibfnamefont {J.}~\bibnamefont {Gacon}}, \bibinfo
  {author} {\bibfnamefont {T.~P.}\ \bibnamefont {Gujarati}}, \bibinfo {author}
  {\bibfnamefont {I.}~\bibnamefont {Hamamura}}, \bibinfo {author}
  {\bibfnamefont {T.}~\bibnamefont {Imamichi}}, \bibinfo {author}
  {\bibfnamefont {C.}~\bibnamefont {Johnson}}, \bibinfo {author} {\bibfnamefont
  {I.}~\bibnamefont {Liepuoniute}}, \bibinfo {author} {\bibfnamefont
  {M.}~\bibnamefont {Motta}}, \bibinfo {author} {\bibfnamefont
  {M.}~\bibnamefont {Rossmannek}}, \bibinfo {author} {\bibfnamefont {T.~L.}\
  \bibnamefont {Scholten}}, \bibinfo {author} {\bibfnamefont {I.}~\bibnamefont
  {Sitdikov}},\ and\ \bibinfo {author} {\bibfnamefont {S.}~\bibnamefont
  {Woerner}},\ }\href@noop {} {\bibinfo {title} {Entanglement forging
  module}},\ \bibinfo {howpublished}
  {\url{https://github.com/qiskit-community/prototype-entanglement-forging}}
  (\bibinfo {year} {2021})\BibitemShut {NoStop}%
\bibitem [{\citenamefont {Zhu}\ \emph {et~al.}(1997)\citenamefont {Zhu},
  \citenamefont {Byrd}, \citenamefont {Lu},\ and\ \citenamefont
  {Nocedal}}]{zhu1997algorithm}%
  \BibitemOpen
  \bibfield  {author} {\bibinfo {author} {\bibfnamefont {C.}~\bibnamefont
  {Zhu}}, \bibinfo {author} {\bibfnamefont {R.~H.}\ \bibnamefont {Byrd}},
  \bibinfo {author} {\bibfnamefont {P.}~\bibnamefont {Lu}},\ and\ \bibinfo
  {author} {\bibfnamefont {J.}~\bibnamefont {Nocedal}},\ }\href
  {https://doi.org/10.1145/279232.279236} {\bibfield  {journal} {\bibinfo
  {journal} {ACM Trans. Math. Softw.}\ }\textbf {\bibinfo {volume} {23}},\
  \bibinfo {pages} {550–560} (\bibinfo {year} {1997})}\BibitemShut {NoStop}%
\bibitem [{\citenamefont {Morales}\ and\ \citenamefont
  {Nocedal}(2011)}]{morales2011remark}%
  \BibitemOpen
  \bibfield  {author} {\bibinfo {author} {\bibfnamefont {J.~L.}\ \bibnamefont
  {Morales}}\ and\ \bibinfo {author} {\bibfnamefont {J.}~\bibnamefont
  {Nocedal}},\ }\href {https://doi.org/10.1145/2049662.2049669} {\bibfield
  {journal} {\bibinfo  {journal} {ACM Trans. Math. Softw.}\ }\textbf {\bibinfo
  {volume} {38}},\ \bibinfo {pages} {7} (\bibinfo {year} {2011})}\BibitemShut
  {NoStop}%
\bibitem [{\citenamefont {Mizukami}\ \emph {et~al.}(2020)\citenamefont
  {Mizukami}, \citenamefont {Mitarai}, \citenamefont {Nakagawa}, \citenamefont
  {Yamamoto}, \citenamefont {Yan},\ and\ \citenamefont
  {Ohnishi}}]{mizukami2020orbital}%
  \BibitemOpen
  \bibfield  {author} {\bibinfo {author} {\bibfnamefont {W.}~\bibnamefont
  {Mizukami}}, \bibinfo {author} {\bibfnamefont {K.}~\bibnamefont {Mitarai}},
  \bibinfo {author} {\bibfnamefont {Y.~O.}\ \bibnamefont {Nakagawa}}, \bibinfo
  {author} {\bibfnamefont {T.}~\bibnamefont {Yamamoto}}, \bibinfo {author}
  {\bibfnamefont {T.}~\bibnamefont {Yan}},\ and\ \bibinfo {author}
  {\bibfnamefont {Y.-y.}\ \bibnamefont {Ohnishi}},\ }\href
  {https://journals.aps.org/prresearch/abstract/10.1103/PhysRevResearch.2.033421}
  {\bibfield  {journal} {\bibinfo  {journal} {Phys. Rev. Research}\ }\textbf
  {\bibinfo {volume} {2}},\ \bibinfo {pages} {033421} (\bibinfo {year}
  {2020})}\BibitemShut {NoStop}%
\bibitem [{\citenamefont {Sokolov}\ \emph {et~al.}(2020)\citenamefont
  {Sokolov}, \citenamefont {Barkoutsos}, \citenamefont {Ollitrault},
  \citenamefont {Greenberg}, \citenamefont {Rice}, \citenamefont {Pistoia},\
  and\ \citenamefont {Tavernelli}}]{sokolov2020quantum}%
  \BibitemOpen
  \bibfield  {author} {\bibinfo {author} {\bibfnamefont {I.~O.}\ \bibnamefont
  {Sokolov}}, \bibinfo {author} {\bibfnamefont {P.~K.}\ \bibnamefont
  {Barkoutsos}}, \bibinfo {author} {\bibfnamefont {P.~J.}\ \bibnamefont
  {Ollitrault}}, \bibinfo {author} {\bibfnamefont {D.}~\bibnamefont
  {Greenberg}}, \bibinfo {author} {\bibfnamefont {J.}~\bibnamefont {Rice}},
  \bibinfo {author} {\bibfnamefont {M.}~\bibnamefont {Pistoia}},\ and\ \bibinfo
  {author} {\bibfnamefont {I.}~\bibnamefont {Tavernelli}},\ }\href
  {https://aip.scitation.org/doi/10.1063/1.5141835} {\bibfield  {journal}
  {\bibinfo  {journal} {J. Chem. Phys}\ }\textbf {\bibinfo {volume} {152}},\
  \bibinfo {pages} {124107} (\bibinfo {year} {2020})}\BibitemShut {NoStop}%
\bibitem [{\citenamefont {Lee}\ \emph {et~al.}(2021)\citenamefont {Lee},
  \citenamefont {Malone}, \citenamefont {Morales},\ and\ \citenamefont
  {Reichman}}]{lee2021spectral}%
  \BibitemOpen
  \bibfield  {author} {\bibinfo {author} {\bibfnamefont {J.}~\bibnamefont
  {Lee}}, \bibinfo {author} {\bibfnamefont {F.~D.}\ \bibnamefont {Malone}},
  \bibinfo {author} {\bibfnamefont {M.~A.}\ \bibnamefont {Morales}},\ and\
  \bibinfo {author} {\bibfnamefont {D.~R.}\ \bibnamefont {Reichman}},\ }\href
  {https://pubs.acs.org/doi/10.1021/acs.jctc.1c00100} {\bibfield  {journal}
  {\bibinfo  {journal} {J. Chem. Theory Comput}\ }\textbf {\bibinfo {volume}
  {17}},\ \bibinfo {pages} {3372} (\bibinfo {year} {2021})}\BibitemShut
  {NoStop}%
\bibitem [{\citenamefont {Blunt}\ \emph {et~al.}(2018)\citenamefont {Blunt},
  \citenamefont {Alavi},\ and\ \citenamefont {Booth}}]{blunt2018nonlinear}%
  \BibitemOpen
  \bibfield  {author} {\bibinfo {author} {\bibfnamefont {N.~S.}\ \bibnamefont
  {Blunt}}, \bibinfo {author} {\bibfnamefont {A.}~\bibnamefont {Alavi}},\ and\
  \bibinfo {author} {\bibfnamefont {G.~H.}\ \bibnamefont {Booth}},\ }\href
  {https://journals.aps.org/prb/abstract/10.1103/PhysRevB.98.085118} {\bibfield
   {journal} {\bibinfo  {journal} {Phys. Rev. B}\ }\textbf {\bibinfo {volume}
  {98}},\ \bibinfo {pages} {085118} (\bibinfo {year} {2018})}\BibitemShut
  {NoStop}%
\end{thebibliography}%

\end{document}